\documentclass[a4paper,11pt]{article}
\usepackage{jheppub} 
\usepackage{lineno}

\usepackage{booktabs}
\usepackage{mathrsfs}
\usepackage{amsmath,amssymb}
\usepackage{dsfont}
\usepackage{xcolor}
\usepackage{comment}
\usepackage{tikz}
\usepackage[normalem]{ulem}

\newcommand{\bea}{\begin{eqnarray}}
\newcommand{\eea}{\end{eqnarray}}

\newcommand{\NC}{{\cal N}_{\CC}}
\newcommand{\NL}{{\cal N}_{\LL}}

\newcommand{\Z}{{\mathbb{Z}}}
\newcommand{\M}{{{M}}}
\newcommand{\R}{{\mathbb{R}}}

\newcommand{\LL}{{{\cal L}}}
\newcommand{\CC}{{{\cal C}}}

\newcommand{\SSS}{{\mathbb{S}}}

\newcommand{\g}{{\mathcal{G}}}

\newcommand{\HH}{{{\cal H }}}

\newcommand\Sp{{\rm Sp}}
\newcommand\Or{{\rm O}}
\newcommand\SO{{\rm SO}}
\newcommand\SU{su}
\newcommand\U{{\rm U}}

\title{\boldmath Holographic description of 4d Maxwell theories and their code-based ensembles}

\author{Ahmed Barbar,}
\author{Anatoly Dymarsky}
\author{and Alfred Shapere}
\affiliation{Department of Physics and Astronomy, University of Kentucky,\\ 506 Library drive, Lexington, KY 40506}

\emailAdd{a.barbar@uky.edu}
\emailAdd{a.dymarsky@uky.edu}
\emailAdd{shapere@g.uky.edu}

\abstract{We formulate a precise holographic duality between an ensemble of 4d $\U (1)^g$ Maxwell theories living on a spin four-manifold $\M_4$ and an Abelian BF-type 2-form gauge theory of level $N$, summed over all five-manifolds with boundary $\M_4$.  The elements of the boundary ensemble are Abelian gauge theories specified by self-dual symplectic codes over $\Z_N$, that parameterize topological boundary conditions in the 5d TQFT. 
Similarly, the equivalence classes of topologies distinguished by the 5d theory are parameterized by orthogonal self-dual codes. Hence the holographic duality can be reformulated in the language of quantum stabilizer codes.  
This duality is closely related to the holographic relationship between ensembles of Narain conformal field theories in 2d and level-$N$ Abelian Chern-Simons theories in 3d. 
In both contexts, the duality extends to correlation functions. 
In the large-$N$ limit, we find that the boundary ensemble average converges to an integral over the moduli space of the gauge couplings and, when finite, is equal to an Eisenstein series of the orthogonal group, a version of the Siegel-Weil formula that  appears in the 2d/3d context.  As a spinoff, we clarify the holographic relationship between the gauge group of the 4d ${\cal N} = 4$ super Yang-Mills theory and the boundary conditions of the  singleton sector in the bulk. 
}

\begin{document}
\maketitle
\flushbottom

\section{Introduction}
\label{sec:intro}

Nearly three decades after the emergence of holography as a fundamental framework for string theory and quantum gravity,  there are still many aspects of the holographic correspondence that are incompletely understood.  In particular, over the past few years it has been realized that the paradigm of holography as a relation between a single quantum field theory and a gravitational theory requires modification.  In a variety of low-dimensional examples \cite{Saad:2019lba,Afkhami-Jeddi:2020ezh,Maloney:2020nni}, it has been found that gravitational theories are dual not to a single quantum field theory, but to a whole ensemble of QFTs.  

In this paper we will establish a new class of holographic dualities between ensembles of four-dimensional Abelian gauge theories and five-dimensional topological gravity theories.  We will relate ensemble averages of U$(1)^g$ Maxwell theories living on a closed 4-manifold ${\cal M}_4$ to five-dimensional topological quantum field theories (TQFTs), summed over 5-manifolds that end on  ${\cal M}_4$.  The TQFTs in question are 5d BF-type Chern-Simons theories involving 2-form gauge fields $B_2$ and $C_2$.



The composition of the ensemble and the set of inequivalent topologies appearing in the sum are controlled by the level $N$ of the Chern-Simons theory.  At level $N=1$, the ensemble consists of a single boundary theory -- as in conventional holography -- and the sum over bulk 5-manifold topologies is trivial, because the bulk theory then 
does not distinguish between topologies.  For $N>1$, the ensembles that arise can be described in terms of codes. These are self-dual symplectic codes of length $2g$ over $\Z_N$ that parametrize  topological boundary conditions of the 5d theory. Each 4d partition function in the ensemble corresponds to a stabilizer state, i.e.~a quantum code defined in terms of the symplectic one. Likewise, the set of equivalence classes of topologies that are distinguished by the bulk CS theory are associated with self-dual orthogonal codes and corresponding stabilizer states. 

We can schematically represent our holographic relation  as follows: 
\begin{equation}\label{schematic}
\langle Z_{\rm Maxwell}\rangle = \sum_{\rm topologies} Z_{\rm 5d\, CS}    
\end{equation}
where the left side is the ensemble averaged partition function
and the $Z_{\rm 5d\, CS}$ on the right is a wavefunction of the bulk CS theory on a topology with a fixed boundary. This holographic duality extends to include correlation functions of $\U (1)$ primaries.  Each partition function appearing in the boundary ensemble is  invariant under the 4d modular group $\Or(n,n,\Z)$.  
There is an additional group $\Sp (2g,\Z_N)$ inherited from the bulk, where it is a 0-form symmetry, that maps theories in the ensemble to each other. 
This is not the S-duality group, which is a symmetry of each 4d theory. On the right-hand side, each bulk wavefunction is invariant under  $\Sp (2g,\Z_N)$, while the set of 5d topologies decomposes into orbits of $\Or(n,n,\Z)$. The combined invariance under both $\Sp (2g,\Z_N)$ and $\Or(n,n,\Z)$  when the level $N$ is square-free constrains both sides  to be equal up to normalization, which can be fixed on physical grounds \cite{Dymarsky:2025agh}.

Our setup and results have closely related analogues in 2d/3d, where ensembles of Narain conformal field theories (CFTs) are known to be dual to Abelian Chern-Simons theories.  In that context, it has been established that the CFT partition function, averaged over a discrete set of Narain moduli specified by codes, is reproduced by the path integral of a  Chern-Simons theory, summed over all possible topologies with a fixed boundary \cite{Aharony:2023zit,Dymarsky:2025agh}.  
It has been known for a long time that the partition function of the 2d Narain CFT for worldsheet modulus $\Omega_{ij}$ and Narain modulus $E_{IJ}$ is closely related to the partition function of 4d U$(1)^g$ Maxwell with coupling matrix $\Omega_{ij}$ on a 4-manifold with modulus $E_{IJ}$ \cite{witten1995s,verlinde1995global}. In the 4d Maxwell case, the spacetime and target space modular groups, respectively $\Or(n,n,\Z)$ and $\Sp(2g,\Z)$, play roles that are reversed relative to their roles in Narain CFTs. Thus the elements of the Maxwell ensemble, parameterized by symplectic codes, correspond to equivalence classes of topologies in the 3d CS theory. Similarly,  distinct Narain theories in the ensemble,  specified by orthogonal codes, correspond to equivalence classes of the 5d topologies appearing in the sum on the right side of \eqref{schematic}. 
The origin of this Narain-Maxwell duality lies in 6d/7d, the duality between the 6d Abelian self-dual two-form gauge theory on $\Sigma\times {\cal M}_4$ and the 7d  three-form CS theory \cite{verlinde1995global,witten2004conformal}.  

It is especially interesting to consider the large-level limit of the Maxwell ensemble.  As $N\to \infty$, the ensemble-averaged  partition function becomes an integral over the entire moduli space of U$(1)^g$ gauge couplings (which is identical to the fundamental domain of genus-$g$ Riemann surfaces), and the sum over 5d topologies approaches a sum over 5d handlebodies, yielding an Eisenstein series of the orthogonal group.  This is similar to 2d/3d result, where the large-$N$ limit yields 
the Siegel-Weil formula, and reproduces the results of  \cite{Afkhami-Jeddi:2020ezh,Maloney:2020nni}. 

Thanks to the exact solvability of both sides of our duality, we will be able to address some central questions about ensemble holography:   When does an ensemble of boundary theories have a holographic dual?  How are the ensemble weights determined?  Which topologies should be summed over in the bulk theory and with what weights?  

Addressing these questions helped us clarify the conventional holographic dictionary of ${\cal N} = 4$ SYM  dual to Type IIB supergravity. It is well known that the bulk includes a topological sector of the kind we discuss in this paper \cite{Witten:1998wy}. We will see that different boundary conditions for the  TQFT fields correspond to different theories at the boundary: $\U(N)$ SYM theory or one of the ${\rm SU}(N)$ theories. 



The paper is organized as follows.
In Section 2, we carry out the holomorphic quantization of 5d Abelian Chern-Simons theory on  $\M_4\times \R$, where $\M_4$ is a closed spin four-manifold. In section 3, we discuss the relation to SymTFT  and the connection to codes.  In section 4, we discuss the holographic duality between ensembles of Maxwell theories and the 5d Chern-Simons theory summed over topologies. Section 5 reconsiders the conventional duality between ${\cal N}=4$  SYM and the type IIB string on $AdS_5\times S^5$, and describe the relation between the boundary conditions of the TQFT  and the gauge group.  We conclude in Section 6 with a discussion of how ensemble duality can be extended to more complex cases involving a dynamical graviton.  Some derivations and detailed calculations have been relegated to a series of appendices. 

\section{Quantization of 5d BC theory}

In this section we construct the Hilbert space ${\cal H}_{\M_4}$ of  topological five-dimensional theory
\bea
\label{5d}
{N\over 2\pi} \sum_{I=1}^g \int B_2^I \wedge dC_2^I
\eea
quantized on  $\M_4\times \R$, where $\M_4$ is a Euclidean orientable spin four-manifold. We first notice that ${\cal H}_{\M_4}$ depends only on the 2-cohomology structure of $\M_4$; hence we can morph $\M_4$ into  $\M_4'$ with trivial odd cohomology while preserving the even cohomology structure, such that  ${\cal H}_{\M_4}$ and ${\cal H}_{\M_4'}$ are isomorphic. Thus, in what follows we assume $\M_4$ has trivial $H^1(\M_4,\Z)$ and $H^3(\M_4,\Z)$.  With this choice of $\M_4$, the theory \eqref{5d} is a dimensional reduction of the seven-dimensional Chern-Simons-type theory
\bea
\label{7d}
{N\over 4\pi} \int H_3 \wedge dH_3 
\eea
compactified on a Riemann surface $\Sigma$ of genus $g$. Alternatively, \eqref{7d} compactified on $\M_4$ is a conventional three-dimensional Chern-Simons theory
\bea
\label{3d}
{K_{ij}\over 4\pi} \int A^i \wedge dA^j,
\eea
with $K_{ij}=N\, \eta_{ij}$, where $\eta$ is the integer-valued intersection form on $H^2(\M_4,\Z)$.
 Since by assumption $\M_4$ is spin, $\eta$ is even and the 3d theory \eqref{3d} is bosonic. 
As was pointed out in \cite{verlinde1995global}, the Hilbert space $\HH_{\M_4}$ of \eqref{5d} on $\M_4\times \R$ will be isomorphic to the Hilbert 
space $\HH_\Sigma$ of \eqref{3d} on $\Sigma\times \R$, both being subspaces of the  Hilbert space of \eqref{7d} on $\M_6\times \R$, where $\M_6=\Sigma \times \M_4$. In what follows we will simply refer to the Hilbert  space $\HH_{\M_4}\simeq\HH_\Sigma$ as $\HH$.

Our main focus will be on holography, in the context of which $\M_4$ will be the boundary of a five-dimensional bulk manifold $X_5$. Such an $\M_4$ must have signature $\sigma=b_2^+ -b_2^-=0$,\footnote{This is because such an $X_5$ would be a cobordism between $M_4$ and the empty set (which has signature 0), and the signature is a complete cobordism invariant of spin 4-manifolds.} in which case $\eta$ can always be brought to the form 
\begin{eqnarray}
\label{eta}
\eta=\left(\begin{array}{cc}
0 & \mathbb{1}_n\\
\mathbb{1}_n & 0
\end{array}\right),\qquad n=b_2^+=b_2^-.
\end{eqnarray}
We will refer to $n$ as the ``genus'' of $\M_4$.

For a closed, oriented, simply-connected Euclidean  spin 4-manifold $\M_4$, Freedman's theorem \cite{freedman1982topology}  then implies that $\M_4$ is homeomorphic to $\#^n\,\SSS^2\times \SSS^2$, the connected sum of $n$ copies of $\SSS^2\times \SSS^2$.     

To describe the Hilbert space of the theory \eqref{5d} on $\M_4\times \R$, it will be convenient to make use of the equivalence of $\HH_{\M_4}$ and $\HH_\Sigma$. From \eqref{eta},
the Chern-Simons theory \eqref{3d}  resulting from the compactification of the 7d theory \eqref{7d} on $\M_4$ takes the form 
\bea
\label{AB}
{N\over 4\pi}\sum_{i=1}^n \int A^i \wedge dB^i+B^i \wedge dA^i.
\eea
This theory was  recently studied in detail in \cite{Aharony:2023zit,Dymarsky:2025agh}. A basis for ${\cal H}_\Sigma$, and hence for ${\cal H}_{\M_4}$, can be defined by choosing a handlebody $X_3$ ending on $\partial X_3=\Sigma$. The basis states
\bea
\label{3dbasis}
|(\alpha,\beta)\rangle_{3d},\qquad \alpha^i_I,\beta^i_I \in \Z_N, 
\eea
are given by the 3d path integral on $X_3$ with Wilson lines of $A^i$ and $B^i$ wrapping the non-shrinkable cycles of $X_3$ $\alpha^i_I$ and $\beta^i_I$ times correspondingly.  
The subscript 3d in \eqref{3dbasis} serves as a reminder that this basis for $\HH$ has a natural origin in 3 dimensions. 

The 3d Chern-Simons theory \eqref{AB} has an explicit $\Or(n,n,\Z_N)$  symmetry that acts on the gauge fields $A^i$ and $B^i$ while preserving the quadratic form $K_{ij}=N\eta_{ij}$. The  action of $\Or(n,n,\Z_N)$ on $\HH$  is implemented by surface operators, and is straightforward in the basis \eqref{3dbasis}: an element $h\in \Or(n,n,\Z_N)$ acts via a unitary operator 
\bea
\label{orthgroupaction}
U_{h}|(\alpha,\beta)\rangle_{3d}=|h (\alpha,\beta)\rangle_{3d},
\eea
where on the right side $h$ acts on $(\alpha^i,\beta^i)$ as a fundamental vector mod $N$.  The action of $\Sp(2g,\Z_N)$  on \eqref{3dbasis}, which is derived from the representation of the mapping class group $\Sp(2g,\Z)$ on $\Sigma$, is more involved and can be found in Appendix \ref{Modular}.

From the point of view of the 5d theory, the basis \eqref{3dbasis} simplifies  the action of the mapping class group of $\M_4$ while the action of the symmetry group of the bulk theory $\Sp(2g,\Z_N)$ is convoluted. (If the original  $\M_4$ is not connected or has nontrivial odd cohomology, its mapping class group is a subgroup of $\Or(n,n,\Z)$. Nevertheless since $\HH$ only depends on the 2-cohomology of $M_4$, the action of the whole $\Or(n,n,\Z)$ is well-defined.)
To make the $\Sp(2g,\Z_N)$ symmetry of the 5d theory \eqref{5d} manifest we introduce a different basis for $\HH$,
\bea
\label{5dbasis}
|(a,b)\rangle_{5d}={1\over N^{gn/2}}\sum_{\alpha,\beta\in \Z_N^{gn}}  e^{{2\pi i\over N}{\rm Tr}(a\,\alpha)}\,\delta_{b,\beta^T}\,|(\alpha,\beta)  \rangle_{3d}, \qquad a^I_i,b^I_i\in \Z_N.
\eea
One can check straightforwardly that the action of $\gamma\in \Sp(2g,\Z_N)$ on \eqref{5dbasis} is analogous to \eqref{orthgroupaction},
\bea
\label{simplgroupaction}
U_\gamma |(a,b)\rangle_{5d}=|\gamma (a,b)\rangle_{5d},
\eea
where $\gamma$ acts on $(a^I,b^I)$ as a fundamental vector. This action is realized by 4d surface operators of the 5d theory on $\M_4 \times \R$. The price to pay for the simplicity of \eqref{simplgroupaction} is the convoluted form of the transformation of \eqref{5dbasis} under $\Or(n,n,\Z_N)$, which can be found in Appendix \ref{Modular}.

The subscript $5d$ in \eqref{5dbasis} indicates that the corresponding states have a natural interpretation in  5d  as path integrals of \eqref{5d} over a five-dimensional ``handlebody'' $X_5$ with the surface operators of $B^I_2,C^I_2$ wrapping $n$ non-shrinkable 2-cycles $a^I_i$ and $b^I_i$ times correspondingly. By the ``handlebody'' $X_5$ here we mean the geometry 
 homeomorphic to a connected sum of $n$ copies of $B_3\times S^2$ 
(where $B_3$ is a 3-ball), which is characterized by the contractibility of a maximal set of $n$ nonintersecting 2-cycles of $\M_4$. 




The basis \eqref{3dbasis} is well-defined for spin $\M_4$ of any signature $\sigma$, in which case  $\alpha\in \Z_N^n$ and $\beta\in \Z_N^{\bar n}$, where $n\equiv b_2^+$, ${\bar n}\equiv b_2^-$, such that $(\alpha,\beta)$ labels an element of $H_2(\M_4,\Z_N)$. This basis was used to quantize the 5d theory \eqref{5d} in \cite{belov2004conformal}.

In fact the basis \eqref{5dbasis}  has a natural interpretation in 3d as well. It is the basis of states with a fixed holonomy of the gauge filed $A^i$ over $2g$ cycles of $\Sigma$, which means these states diagonalize the action of the Wilson loops of $A^i$.  In the SymTFT context, this is the basis of distinct twisted partition functions of a given boundary theory \cite{Gaiotto:2020iye}. 
Similarly, the basis \eqref{3d} in 5d is the basis of states with fixed $B_2$ flux  over all 2-cycles of $\M_4$. Its SymTFT interpretation is also as a set of twisted partition functions of the boundary 4d theory.


\subsection{Holomorphic quantization}
\label{sec:holquantization}
In the previous section we discussed the structure of the Hilbert space $\HH$ without specifying the explicit form of the wavefunctions. The latter depends on the boundary conditions at $\Sigma$ (in the 3d theory) or $\M_4$ (in the 5d theory). Below we follow the standard holomorphic quantization of the 3d Chern-Simons theory \cite{Bos:1989wa,Elitzur:1989nr} and generalize it to higher $d$. The idea is to start with the 7d theory \eqref{7d}, which is very similar to chiral CS theory in 3d, and quantize it first. Then the  wavefunctions of the 3d theory \eqref{AB} and the 5d theory \eqref{5d} can be obtained by dimensional reduction. 
More details can be found in Appendix \ref{app:dimreduction}.

We begin with the 7d theory \eqref{7d} on $\M_6\times \R$ in a gauge where all components of the 3-form field $H_3$ along $\R$ vanish. In this gauge 
one can expand $H_3$ into harmonic modes and cohomologically trivial fluctuations, generalizing the 3d case \cite{Bos:1989wa},
\bea
H_3={i \pi\over N} \sum_{A,B} \zeta_A  ({\bf \Omega}_2^{-1})^{AB}\omega^{(3)}_B +{\rm c.c.}+\partial \chi.  
\eea
Here $\omega^{(3)}_A$ is a basis of self-dual 3-forms on $\M_6$, $\star\, \omega^{(3)}_A=i\, \omega^{(3)}_A$ and $\bf \Omega$ is the corresponding ``modular'' parameter of the 6d manifold. 
Following \cite{Aharony:2023zit} we add the boundary term to the 7d action \eqref{7d}
\bea
\label{boundaryterm}
{N\over 4\pi}\int_{\M_6} H_3\wedge \star H_3.
\eea
With this boundary term it is consistent to fix the value of $\zeta_A$ at the boundary while allowing its complex conjugate $(\zeta_A)^*$ to fluctuate freely. The resulting wavefunction, which we construct explicitly in Appendix \ref{app:dimreduction}, is a holomorphic function of $\zeta_A$.

Next we consider $\M_6=\Sigma\times \M_4$, with $\Sigma$ and $\M_4$ as above.  There are a total of $2ng$ independent self-dual  three-forms on $\M_6$, see Appendix \ref{app:dimreduction},
\bea
\omega^{}_I \wedge \omega^{+}_i,\qquad \omega^*_I \wedge \omega^{-}_i,\qquad I=1\dots g,\,\, i=1\dots n.
\eea
In this case the vector $\zeta_A$ is  a combination of two holomorphic variables $\xi^i_I$ and $\bar \xi^i_I$ such that 
\bea
\label{H3parametrization}
H_3={i \pi\over \sqrt{N}} \sum_{I,i} \xi_I^i (\Omega_2^{-1})^{IJ} \omega^*_J\wedge \omega^+_i-{i\pi \over \sqrt{N}}  \sum_{I,i} \bar\xi_I^i (\Omega_2^{-1})^{IJ} \omega_J \wedge\omega^-_i+ {\rm c.c.}+\partial \chi.
\eea
At the quantum level the complex conjugate variables $\xi^*$ and $\bar \xi^*$ become operators 
canonically conjugate to $\xi$ and $\bar \xi$ with respect to the measure $e^{-\pi\, {\Omega_2^{-1}}(|\xi|^2+|\bar \xi|^2) }$ inherited from \eqref{boundaryterm} such that,
\bea
\label{diffop}
\xi^* \rightarrow {\Omega_2\over \pi} {\partial \over \partial \xi},\qquad \bar\xi^* \rightarrow {\Omega_2\over \pi}{\partial \over \partial \bar\xi}.
\eea

To obtain wavefunctions of the 3d and 5d theories, one can start with the wavefunction in  7d and then dimensionally reduce it. Alternatively one can start directly in 3d or 5d and add  corresponding boundary terms to \eqref{AB} or  \eqref{5d}. The holomorphic variables $\xi,\bar \xi$  will then emerge parameterizing the harmonic expansions of the 3d and 5d fields. Choosing a gauge in which all components along the non-compact direction vanish,  we have in 3d 
\bea\label{ABxibarxi}
A^i&=&{i\pi (G^{-1/2})_{ij}\over \sqrt{2N}} \left(   \xi^j_I (\Omega_2^{-1})^{IJ}\omega^*_J-  \bar\xi^j_I (\Omega_2)^{-1}_{IJ}\omega^{}_J\right)+{\rm c.c.}+\partial \chi_A,\\
\nonumber
B^i&=&{i\pi (G^{-1/2})_{ij}\over \sqrt{2N}}\left( (G-B)_{jk}  \xi^k_I (\Omega_2)^{-1}_{IJ}\omega^*_J+(G+B)_{jk}  \bar\xi^j_I (\Omega_2)^{-1}_{IJ}\omega^{}_J\right)+{\rm c.c.}+\partial \chi_B,
\eea
and the fluctuating part $\partial \chi_{A,B}$ vanishes at the boundary. 
 Similarly, in 5d
\bea
\label{BCxibarxi}
&&B_2^I ={i\pi (\Omega_2^{-1})^{IJ}\over \sqrt{N}}\left(\xi_J^i \omega_i^+-\bar \xi_J^i \omega_i^-\right)+{\rm c.c.}+\partial \chi_B,\\ \nonumber
&&C_2^I ={i\pi (\Omega_2^{-1})^{JK}\over \sqrt{N}}\left(\Omega^*_{IJ}\xi_K^i \omega_i^+-\Omega_{IJ}\bar \xi_K^i \omega_i^-\right)+{\rm c.c.}+\partial \chi_C.
\eea

There is another, equivalent way to construct the wavefunctions. The Hilbert space $\HH$ can be defined as a representation of the group of surface operators wrapping the three-cycles $\Gamma$ in $\M_6$,
\bea
\label{SO}
W_{\Gamma({\rm n})}={\rm exp} \{2\pi i\int_{\Gamma} H_3\},\qquad \Gamma^{\vee}=\sum_{I,i}{\rm n}^{i}_I\,\omega^{(1)}_I\wedge \omega_i^{(2)},\qquad {\rm n}\in \Z^{4gn}_N.
\eea
Upon compactification on $\M_4$, in  3d  these operators become Wilson lines of the $A$ and $B$ gauge fields wrapping one-cycles of $\Sigma$, while after compactifying on $\Sigma$, in  5d they became the surface operators of $B_2$ and $C_2$ wrapping two-cycles of $\M_4$. 
Mathematically, $W_{\Gamma}$ are holomorphic differential operators of $\xi,\bar\xi$. 
Multiplication by $\xi,\bar\xi$ together with  \eqref{diffop} form the Heisenberg algebra, while 
the surface operators \eqref{SO} form a Heisenberg group
\bea
\label{Heisenberg}
W_{\Gamma}W_{\Gamma'}=W_{\Gamma+\Gamma'} e^{{i\pi \over N} \left(\Gamma \cap \Gamma'\right)},\quad \left(\Gamma \cap \Gamma'\right)=-{\rm Tr}({\rm n}^T \eta\, {\rm n}' J). 
\eea
Here ${\rm n,n}'$ are $2n$ by $2g$ matrices. 
The Stone-von Neumann theorem  implies that such a  representation is unique \cite{mumford2007tata}, which is another way to see  that $\HH_\Sigma$ and $\HH_{\M_4}$ are isomorphic. The wavefunctions forming the representation of \eqref{Heisenberg} can be written explicitly in terms of the theta functions \cite{mumford2007tata,gelca2010classical}, as discussed in Appendix \ref{app:dimreduction}.

The explicit form of the wavefunctions  was recently  reviewed  in \cite{Aharony:2023zit,Dymarsky:2025agh},
\begin{align}
\label{3dtheta}
    & \Theta_{c_1 c_2  \ldots  c_g}(\Omega, E) \ = \ {\rm det}(\Omega_2)^{n/2}\sum_{ u_1,  \ldots   u_g} e^{i \pi {u}^T_I \Pi^{IJ}  {u}_J+2\pi i\, (p^I_L\cdot \xi_I-p^I_R \cdot \bar \xi_I) +\pi(\xi \Omega_2^{-1}\xi^T +{\bar\xi} \Omega_2^{-1}\bar\xi^T)/2}.
\end{align}
Here 
\bea
{u}_I &=& {1\over \sqrt{2}}
\left(\begin{array}{c}
p_L^I+p_R^I\\
p_L^I-p_R^I
\end{array}
\right)= \mathcal{O} ( n_I \sqrt{N}+c_I / \sqrt{N} \, ), \qquad c_I=(\alpha_I,\beta_I) \in \Z_N^{2n},\\
\Pi^{IJ}_{ij}&=&(\Omega_1)_{IJ} \eta_{ij} +i (\Omega_2)_{IJ} \delta_{ij},
\eea
and the sum goes over all $n_I \in \Z^{2n},$.
The orthogonal matrix $\mathcal{O} \in \Or(n, n, \mathbb{R})$ is defined in terms of the ``Narain data'' $E=G+B$,    metric $G$ and $B$-field, specified by the metric on $\M_4$,
\bea
\label{Omatrix}
 &&\mathcal{O}(E)=G^{-1/2}\left(\begin{array}{cc}
 \mathbb{1} &  B\\
 0 & G
 \end{array}\right).
\eea

We stress that the dependence on the metric on $\Sigma$, encoded in $\Omega$, as well as on the metric on $\M_4$, encoded in $E=G+B$, comes from the boundary term  \eqref{boundaryterm} (or its dimensional reduction). The corresponding boundary conditions with $\xi,\bar\xi$ fixed and $\xi^*,\bar\xi^*$ fluctuating define a particular state $\langle \Omega,E,\xi,\bar \xi| \in \HH^*$. The wavefunction  is a matrix element between  this state and a state in $\HH$. An interpretation of this construction in terms of SymTFT will be discussed in the next section. 

The modular $\Sp(2g,\Z)$ and orthogonal $\Or (n,n,\Z)$ groups -- mapping class groups of $\Sigma$ and $\M_4$, that act on $\HH$ by $\Sp(2g,\Z_N)$ and $\Or (n,n,\Z_N)$ discussed above -- act  on the boundary state as follows
\bea
\label{modtrbra}
&& U_\gamma |\Omega,E,\xi,\bar \xi\rangle =|\Omega_\gamma,E,\xi_\gamma,\bar \xi_\gamma\rangle,\\
&&\Omega_\gamma=(A\Omega+B)(C\Omega+D)^{-1},\quad \xi_\gamma=\xi (C\Omega+D)^{-1},\quad \bar\xi_\gamma=\bar\xi (C\Omega^*+D)^{-1}, \label{modulartransformation}
\eea
and 
\bea
&& U_h |\Omega,E,\xi,\bar \xi\rangle =|\Omega,E_h,\xi_h,\bar \xi_h\rangle,\\
&& (E_h)^T=(A E_h^T+B)(CE_h^T+D)^{-1},\quad 
\xi_h=O_L\, \xi, \quad \bar\xi_h=O_R\,\bar\xi,
\label{orthogonaltransformation}
\eea
where orthogonal matrices $(O_L,O_R)\in O(n,\R)\times O(n,\R)= O(2n,\R)\cap O(n,n,\R)$ are defined by  $(O_L\oplus O_R)(h,E)=H{\mathcal O}(E_h)\, h\,{\mathcal O}^{-1}(E)H^{-1}$ and $H$ is the $2\times 2$ Hadamard matrix tensor $\mathbb{1}_n$.

Actual wavefunctions in 3d and 5d  will also include contributions from the fluctuating modes,
\bea
\label{3dblock}
\langle \Omega,E,\xi,\bar \xi|(\alpha,\beta)\rangle_{3d}=\Psi_{c_1\dots c_g}(\Omega,E,\xi,\bar\xi)={\Theta_{c_1 c_2  \ldots  c_g} \over \Phi}.
\eea
The ket states $\langle \Omega,E,\xi,\bar \xi|$ above are different  in 3d and 5d theories by a scalar factor. 
In 3d theory $\Phi$ is determined by a scalar Laplacian on $\Sigma$ \cite{Belov:2005ze,Maloney:2020nni,Porrati:2019knx},
\bea
\label{3dPhi}
\Phi_{3d}=({\rm det}'\Delta_0)^{n/2},
\eea
while in 5d theory  $\Phi$ is determined by the scalar and vector Laplacians on the four-manifold \cite{belov2004conformal}
\bea
\label{5dPhi}
\Phi_{5d}= \left({{\rm det'}\, \Delta_1\over {\rm det}'\, \Delta_0}\right)^{g/2}.
\eea
In both cases $\Phi$  does not carry any quantum charges and is invariant under both orthogonal and symplectic groups (with one group not acting on $\Sigma$ or $\M_4$ at all, while the other merely relabeling the cohomologies). 
For $g=1$, expression in \eqref{3dPhi} simplifies to $\Phi_{3d}=\tau_2^{n/2}|\eta(\tau)|^{2n}$.

A straightforward computation yields for the wavefunctions of the basis \eqref{5dbasis},
\bea
\label{5dblock}
\langle \Omega,E,\xi,\bar \xi|(a,b)\rangle_{5d}={\sf \Psi}_{{\rm c}_1\dots {\rm c}_n}(\Omega,E,\xi,\bar\xi)= 
{{\varTheta}_{{\rm c}_1\dots {\rm c}_n}\over  \Phi}.
\eea
Here  ${\rm c}_i=(a_i,b_i)\in \Z^{2g}$ and 
\bea
\label{5dtheta}
\varTheta_{{\rm c}_1\dots {\rm c}_g}&=&{\rm det}(G)^{g/2}\sum_{{\rm v}_1\dots {\rm v}_n }e^{- \pi |{\rm v}|^2-2\pi i \tilde{n}^T B\tilde{m}+\sqrt{2}\pi(\bar\xi \Omega_2^{-1/2} {\rm v}-\xi  \Omega_2^{-1/2} {\rm v}^*)+\pi\, \bar \xi\, \Omega_2^{-1}\xi^T},\\
{\rm v}&=&G^{1/2}\Omega_2^{-1/2}(\tilde{n}+\Omega \tilde{m}),\quad \tilde{n}_i=(n_i N+a_i)/\sqrt{N},\quad \tilde{m}_i=(m_i N+b_i)/\sqrt{N},\quad n_i,m_i, a_i,b_i\in \Z_N^{g}.\nonumber
\eea


To conclude this section, we discuss the relation between the holomorphic quantization discussed above and the quantization scheme discussed in the appendix A of \cite{Maldacena:2001ss}.
The $D$-dimensional spacetime there is assumed to be a cylinder $X_D=M_{D-1}\times \R$ with Minkowski signature.  
Starting from the BF-type Abelian theory
\bea
\label{AAaction}
{N\over 2\pi}\int A_{p+1}\wedge dA_{D-p-2},
\eea
and following \cite{Maldacena:2001ss} we add the boundary term on $\partial M_{D-1}\times \R$,
\bea\label{bt}
{1\over 4r^2}\int A_{D-p-2}\wedge \star A_{D-p-2},
\eea
which yields the self-dual boundary condition
\bea
\label{selfdual}
{Nr^2\over \pi}A_{p+1}=\star A_{D-p-2}.
\eea
As we will show momentarily this is the same as the holomorphic quantization described above, in the particular case of vanishing $\xi=\bar\xi$. To see that, we first note that the near-boundary region $\partial M_{D-1}\times \R$ is related to $\Sigma\times \R$ for the 3d theory  (or $\M_4\times \R$ for the 5d theory) after the Wick rotation. 
Next, the 
holomorphic quantization assumes that $\xi,\bar\xi$ are fixed at the boundary, while their complex conjugates $\xi^*,\bar\xi^*$ fluctuate freely. Focusing on the 3d case and ignoring for now that this condition renders $A,B$ complex, by taking $\xi=\bar\xi=0$  we find from \eqref{ABxibarxi}
\bea
B^i=i G_{ij} \star A^j-(G^{-1/2}BG^{1/2})_{ij} A^j.
\eea
After taking $n=1$, such that $G=r^2$ and $B=0$ and rotating to Minkowski signature we recover \eqref{selfdual} after identifying 
\bea
A_{p+1}=\sqrt{\pi\over N}A,\qquad 
A_{D-p-2}=\sqrt{N\over \pi}B,\qquad D=3,\quad p=0.
\eea
At this point we can match the boundary term \eqref{bt}  to be equal on-shell to the difference between \eqref{AAaction} and \eqref{AB}, while the boundary term \eqref{3dbt} vanishes on-shell.  

In a similar way, the holomorphic quantization of the 5d theory \eqref{5d} with $\xi=\bar\xi=0$, after rotating to Minkowski signature, leads to 
\bea
\star (C_2-\Omega\, B_2)=i(C_2-\Omega\, B_2).
\eea
Focusing on the case with $g=1$, for which we can replace $\Omega\rightarrow \tau$, and taking for simplicity $\tau_1=0$, this reduces to \eqref{selfdual} after the identifications
\bea
A_{p+1}={\sqrt{\pi \tau_2}\over \sqrt{N}r}B_2,\qquad A_{D-p-2}={\sqrt{N}r\over \sqrt{\pi \tau_2}} C_2,\qquad D=5,\quad p=1.
\eea
To summarize, the self-duality boundary condition \eqref{selfdual} is the counterpart of holomorphic quantization when the system is quantized in Minkowski signature on a cylinder, as is done in \cite{Elitzur:1989nr}.
A generalization of \eqref{selfdual} to allow non-zero $\xi,\bar\xi$ was recently discussed  in \cite{Antinucci:2024bcm}.

\section{Abelian TQFTs and codes}
\label{sec:SymTFT}

\subsection{SymTFT, topological boundary conditions, and codes}
\label{sec:SymTFTpartI}

As in the case of the 3d theory \eqref{AB}, which is a SymTFT of global $\Z^n_N$ symmetry in two dimensions, the 5d  topological theory \eqref{5d} is a SymTFT of global $\Z^g_N$ symmetry in four dimensions. Any four-dimensional theory with  global $\Z^g_N$ symmetry 
can be coupled to $B_2$ and $C_2$ fields so that the wavefunctions 
\bea
\label{confblocks}
\langle {\cal B}_{4d}|(a,b)\rangle_{5d},  
\eea
are the conformal blocks of the 4d theory on $\M_4$ (which is assumed to be of signature zero). Here  $\langle {\cal B}_{4d}|$ is the state in the  dual to the Hilbert space $\HH_{\M_4}^*$ of the 5d topological theory, created by the boundary conditions on the $B_2,C_2$ fields coupled to 4d theory. This is completely analogous to the 2d case, where   
\bea
\langle {\cal B}_{2d}|(\alpha,\beta)\rangle_{3d}
\eea
are the conformal blocks of the 2d CFT \cite{Gaiotto:2020iye}. 
The wavefunctions of 3d and 5d theories obtained in previous section using holomorphic quantization \eqref{3dblock} and \eqref{5dblock} can be recognized in this language  to be the conformal blocks of 2d Narain and 4d Maxwell theories correspondingly,\footnote{The conformal blocks are themselves the partition functions of generalized Narain  or  Maxwell theories  defined by non-self-dual lattices, e.g.~ \cite{Ashwinkumar:2021kav,Fliss:2023uiv}. These are relative theories in the sense of \cite{Freed:2012bs}.} 
\bea
\langle {\cal B}_{2d/4d}|=\langle \Omega,E,\xi,\bar\xi|.
\eea
Another convenient 4d example is provided by $\SU (N)$ gauge theory, with center $\Z_N$ coupled to 5d $B_2,C_2$ fields \cite{Witten:1998wy, Apruzzi:2021nmk_symtft,Bergman:2022otk}. 

We are primarily interested in the description of topological boundary conditions. It has been recently shown in \cite{Barbar:2023ncl} that the topological boundary conditions of a 3d Abelian Chern-Simons theory are naturally parameterized by classical even self-dual codes. Below we extend this result to 5d and show that topological boundary conditions of the 5d Abelian theory \eqref{5d} are naturally parameterized by classical symplectic self-dual  codes  over $\Z_N$ of length $2g$. We first recall the 3d story, where topological boundary conditions are specified by a maximal (Lagrangian) non-anomalous subgroup ${\cal C}$ of the one-form symmetry group. For the theory \eqref{AB} these are the subgroups of ``codewords'' $c\in {\cal C}\subset \Z_N^{2n}$ closed under addition mod $N$ and satisfying the condition of ``evenness''  
\bea
\label{even}
\alpha\, \cdot\, \beta=0\, \, {\rm mod}\,\, N,\qquad c=(\alpha,\beta)\in \Z_N^n\times \Z_N^n,
\eea 
and self-duality with respect to inner product  \eqref{eta} \cite{Barbar:2023ncl,Dymarsky:2025agh}. In the language of anyons these conditions ensure that the anyons are all bosons and that the group of anyons is maximal. Trivial mutual braiding of all anyons is guaranteed by these conditions, as follows from the linearity of ${\cal C}$ and \eqref{Heisenberg}. 

Similarly, in 5d, topological boundary conditions are specified by additive subgroups ${\cal L}\subset \Z_n^{2g}$ satisfying self-orthogonality 
\bea
\label{selforth}
a_1\cdot b_2-a_2\cdot b_1=0\,\, {\rm mod}\,\, N,\qquad 
{\rm c}_k=(a_k,b_k)\in {\cal L}\subset \Z_N^n \times \Z_N^n,
\eea
and self-duality ${\LL}={\LL}^{\top}$ with respect to 
\bea
\label{J}
J=\left(\begin{array}{cc}
0 & \mathbb{1}_g\\
-\mathbb{1}_g & 0
\end{array}\right).
\eea
In the language  of anyons the group $\LL$ is a collection of anyons closed under fusion and  with trivial mutual braiding, as follows from self-orthogonality and \eqref{Heisenberg}. The self-duality condition ensures that the group $\LL$ is maximal, i.e.~Lagrangian. 

The conditions on $\CC$ and $\LL$ as outlined above are well-known in the literature \cite{kapustin2011topological,aharony2013reading,gaiotto2015generalized,kaidi2022higher}. What is new here is the interpretation of these groups as classical additive codes, which has several advantages.  First, this approach makes an explicit connection with the extensive  literature on ``code CFTs'' \cite{dolan1996conformal,Dymarsky:2020qom,Dymarsky:2020bps,
Henriksson:2021qkt,Dymarsky:2021xfc, Henriksson:2022dnu,
Henriksson:2022dml,Buican:2021uyp,Angelinos:2022umf,Furuta:2022ykh, Alam:2023qac, Furuta:2023xwl, Aharony:2023zit,Barbar:2023ncl,Kawabata:2023nlt, Kawabata:2023iss,Kawabata:2023rlt,  Buican:2023ehi,Ando:2025hwb,Angelinos:2025mjj,Dymarsky:2025agh}. In addition, classical self-dual codes are well studied and certain results from coding theory have immediate applications to SymTFT. For example, the central and very general result of \cite{Nebe} is that the code-based states 
\bea
\label{Cstate}
|\CC\rangle\equiv \sum_{c\in \CC^g}|c\rangle_{3d}
\eea
and 
\bea
\label{Lstate}
|\LL\rangle\equiv \sum_{{\rm c}\in \LL^n}|{\rm c}\rangle_{5d}
\eea
span an (over-)complete basis of $\Or(n,n,\Z)$- and $\Sp(2g,\Z)$-invariant states in $\HH$. A further advantage is an immediate connection to quantum information theory  arising from the fact that \eqref{Cstate} and \eqref{Lstate} are quantum stabilizer states. We elaborate on this point below. 

To illustrate the connection with codes, we recall that classical self-dual symplectic  codes $\LL$ over $\Z_N$ of length $2g$ define  maximal (i.e.~self-dual) quantum stabilizer codes on $g$ qudits of dimension $N$ \cite{Calderbank:1996aj,selforth}.\footnote{
Starting with the  generalized Pauli matrices  $X$ and $Z$ on the dimension $N$ qudit, up to an overall phase the stabilzier generators are $Z^a X^b$ for each $(a,b)\in \LL\subset \Z_N^g \times \Z_n^g$.
} Consider for example  4d gauge theory with the  gauge group $\SU (2)$. After gauging the center $\Z_2$, one obtains an $\SU(2)/\Z_2$ gauge theory known as $\SO(3)_+$. Subsequently shifting $\theta$ by $\pi$ yields the so-called $\SO(3)_-$ theory. These three theories correspond to three topological boundary conditions of the 5d theory \eqref{5d} with level $N=2$. The corresponding Lagrangian subgroups 
\bea
\LL_0=(a,0),\quad \LL_1=(0,b),\quad \LL_2=(a,a),\quad a,b\in \Z_2,
\eea
exhaust the list of all self-dual symplectic codes of length 2 over $\Z_2$.
This description in terms of the subgroups $\LL_k$ is of course exactly the same as in \cite{aharony2013reading}. Using the interpretation of $\LL_k$ as additive self-dual symplectic codes we readily recognize these as three self-dual stabilizer groups of one qubit, generated by $\sigma_z,\sigma_x,$ and $\sigma_y=-i\sigma_z \sigma_x$ correspondingly. 

The equivalence  between the symplectic self-dual codes $\LL$ and maximal (self-dual) stabilizer codes suggests that  topological boundary conditions defined by $\LL$ might admit an interpretation in terms of quantum information. 
This approach  was taken in \cite{Dymarsky:2020bps,Dymarsky:2020qom,Buican:2021uyp}  for codes $\CC$ when $N=2$ (in this case symmetric \eqref{eta} and symplectic \eqref{J} scalar products are equivalent). While the relation between  $\LL$ (or $\CC$ for $N=2$) and quantum stabilizer codes is mathematically correct and useful, e.g.~it was used in \cite{Dymarsky:2020qom} to classify all inequivalent
codes in terms of graphs, it does not immediately provide a physically motivated interpretation.

To interpret topological boundary conditions in terms of quantum error correcting codes, we note that the states in $\HH$ produced by these boundary conditions are given by \eqref{Cstate} and \eqref{Lstate}. It is well-known that the Hilbert space of the Abelian TQFT is equivalent to the Hilbert space of generalized qudits. Thus, braiding of anyons in Abelian theories results only in Clifford gates \cite{Bloomquist:2018hrp}. That is to say, the line operators of the 3d theory wrapping all possible 1-cycles of $\Sigma$ or surface operators of the 5d theory wrapping all possible 2-cycles of $\M_4$ generate the generalized Pauli group of $2gn$ qudits of dimension $N$. Mathematically, this is equivalent to the statement that the  generators $W_\Gamma$ from \eqref{SO} form the Heisenberg group \eqref{Heisenberg}. 
More specifically, taking $X_3$ to be the handlebody used in 3d theory \eqref{AB} to define 
the basis \eqref{3d}, the line operators wrapping around contractable cycles of $X_5$ are the generalized $Z$ gates while those wrapping non-shrinkable cycles are generalized $X$ gates \cite{Dymarsky:2025agh}. Choosing shrinkable one-cycles of $X_3$ to be the first $n$, denoted as $\omega_I^{(1)}$ in Appendix \eqref{app:dimreduction}, we can write explicitly   
\bea
&&W[c_{\rm a},c_{\rm b}]:=W_{\Gamma},\qquad c_{\rm a}=({\rm n},{\rm m}),\quad 
c_{\rm b}=({\rm p},{\rm q}),\quad {\rm n,m,p,q}\in \Z_N^{ng}\\ \nonumber
&&\Gamma^\vee=\sum {\rm n}_I^i\, \omega_I^{(1)} \wedge \omega_i^{(2)}+
{\rm m}_I^i\, \omega_I^{(1)} \wedge \omega_{n+i}^{(2)}+ 
{\rm p}_I^i\, \omega_{g+I}^{(1)} \wedge \omega_i^{(2)}+
{\rm q}_I^i\, \omega_{g+I}^{(1)} \wedge \omega_{n+i}^{(2)},\\
&& W[c_{\rm a},0]|c\rangle_{3d}=e^{{2\pi i \over N} {\rm Tr}(c_{\rm a}^T \eta c)} |c\rangle_{3d},\\
&& W[0,c_{\rm b}]|c\rangle_{3d}=|c+c_{\rm b}\rangle_{3d}.
\eea
Similarly, in 5d surface operators wrapping shrinkable 2-cycle  of $X_5$ used to define the basis \eqref{5dbasis} are generalized $Z$ gates while those wrapping non-shrinkable 2-cycles are generalized $X$ gates,
\bea
&& W[{\rm c}_\alpha;{\rm c}_\beta]:=W[c_{\rm a},c_{\rm b}],\quad {\rm c}_\alpha=(-{\rm p}^T,{\rm n}^T),\quad {\rm c}_\beta=({\rm m}^T,{\rm q}^T),\\
&& W[{\rm c}_{\alpha};0]|{\rm c}\rangle_{5d}=e^{{2\pi i \over N} {\rm Tr}({\rm c}_{\alpha }^T J {\rm c})} |{\rm c}\rangle_{5d},\\
&& W[0;{\rm c}_{\beta}]|{\rm c}\rangle_{5d}=|{\rm c}+{\rm c}_{\beta}\rangle_{5d}.
\eea
Now we readily recognize that \eqref{Cstate} and \eqref{Lstate} are quantum stabilizer states defined by self-dual quantum stabilizer codes. Furthermore the latter are of the Calderbank–Shor–Steane (CSS) type \cite{calderbank1996good,steane1996multiple}, defined in terms of classical codes $\CC$ and $\LL$ correspondingly. Thus the state \eqref{Lstate} created by the topological boundary condition (maximal anyon condensation) defined by $\LL$ is stabilized by all generalized Pauli group elements of the form 
\bea
\label{stabilizer}
W[{\rm c}_\alpha;{\rm c}_\beta] |\LL\rangle=|\LL\rangle,\qquad {\rm c}_\alpha,{\rm c}_\beta \in \LL.
\eea
This defines the state uniquely (up to normalization).
Interpretation of the states produced by topological boundary conditions in 3d Abelian theory as CSS quantum stabilizer states was already given in \cite{Dymarsky:2024frx,Dymarsky:2025agh}. Here we extend it to 5d.\footnote{A construction of the state $|\LL\rangle$ defined by a topological boundary condition in 5d theory, that is equivalent to \eqref{stabilizer} has previously appeared in \cite{Bergman:2022otk}, but without noting the connection to quantum stabilizer codes.}

The connection between anyon condensation and quantum stabilizer codes can be formulated more broadly, by relaxing the condition of self-duality of the underlying classical codes. Namely, the Hilbert space of a theory obtained via (partial) anyon condensation in an Abelian TQFT is always a quantum stabilizer code of CSS type, parameterized by a self-orthogonal  classical code. The meaning of self-orthogonality, i.e.~the choice of the inner product, depends on the theory. In 3d self-orthogonal means even; in 5d it means symplectic. Furthermore, if the anyons are condensed on a codimension one defect, also known as higher gauging \cite{Roumpedakis:2022aik}, it becomes a projector on the code subspace. 
For example, the surface operator in the 3d theory obtained  by gauging the 1-form symmetry $\CC$ on a 2d surface is a projector of the form 
\bea
{1\over |\CC|}\sum_{c_a,c_b\in \CC} W[c_a,c_b].
\eea
Here $\CC$ is even but not necessarily self-dual.

We conclude by mentioning that topological boundary conditions specified by $\LL$ can be understood in the literal sense in terms of the behavior of $B_2,C_2$ near the boundary \cite{Witten:1998wy}. In particular the simplest symplectic code $\LL_0=(*,0)\subset \Z_N\times \Z_N$, where $*$ stands for any element of $\Z_N$, corresponds to the boundary condition 
\bea
C_2=0,\label{topbc}
\eea
while $B_2$ fluctuates freely (Neumann b.c.). Similarly $\LL_1=(0,*)\subset \Z_N\times \Z_N$ would correspond to $B_2=0$ and $C_2$ fluctuating freely at the boundary (Dirichlet b.c.). We note that these, and other boundary conditions are consistent with action \eqref{5d} without any additional boundary terms. They assume real polarization, which is different from the holomorphic quantization discussed above. Imposing e.g.~$C_2=0$ within the holomorphic quantization by fixing the values of $\xi,\bar\xi$ will not impose topological boundary conditions. To impose topological boundary condition in terms of $\xi,\bar\xi$ one can use the complexness of ket states $\langle \Omega,E,\xi,\bar \xi|$  and integrate the boundary wavefunction over $d\xi,d\bar \xi$ with the kernel $\langle \LL|\Omega,E,\xi,\bar\xi\rangle$.

\subsection{Topology and codes}
\label{sec:topologyandcodes}
States $|\LL\rangle$ and $|\CC\rangle$ were discussed above as those defined by topological boundary conditions in 5d and 3d theories correspondingly. As we show below these state have yet another interpretation. As was discussed in \cite{Dymarsky:2024frx}, up to an overall normalization $|\LL\rangle$ are the states produced by the path integrals of 3d theory \eqref{AB} on a 3d manifolds with a boundary. 
Here we revisit and generalize this result to 5d by showing that $|\CC\rangle$ are the states produced by the 5d path integral on 5d topologies. 


We start with the 3d case. 
In what follows we make reference to genus reduction as a way to construct 3d topologies with a boundary. We refer the reader to \cite{Dymarsky:2024frx} for details. We also discuss ``genus reduction'' in 5d later in section \ref{sec:ensemblehol}.

Without loss of generality we can take $n=1$; an arbitrary $n$ can be restored by considering $n$-th tensor power of the resulting state. 
Up to an overall normalization, the vacuum state of the 3d theory $|0\rangle_{3d}^g$ -- the path integral on a handlebody $X_3$ of genus $g$  -- is the code state $|\LL_0\rangle$ where $\LL_0$ is a self-dual symplectic code,
\bea
|0\rangle_{3d}^g={1\over N^{g/2}}|\LL_0\rangle_{5d},\qquad (a_1,\dots,a_g,0,\dots,0)\in\LL_0,\qquad a_i\in \Z_N. 
\eea
In what follows we skip the subscripts $3d,5d$ for simplicity. 
A modular transformation $\gamma\in Sp(2g,\Z)$ maps this state into another state -- the path integral on another handlebody, 
\bea
U_\gamma|0\rangle^g ={1\over N^{g/2}}|\LL_\gamma\rangle,
\eea
which is also a code state associated with another self-dual symplectic code $\LL_\gamma$, where
\bea
{\rm c}=(a_1,\dots,a_g,b_1,\dots,b_g)\in \LL_\gamma,\quad {\rm iff}\quad  {\rm c}=\gamma(a_1',\dots,a'_g,0,\dots,0)\,\, {\rm mod}\,\, N,
\eea
for arbitrary $a'_i\in \Z_N$. After the genus reduction the resulting state 
\bea
\label{resultingstate}
{}^{\tilde{g}}\langle 0|U_\gamma|0\rangle^g=N^{-(g+\tilde{g})/2}\sum_{{\rm x}\in L} m({\rm x})|{\rm x}\rangle \in \HH^{g-\tilde{g}}
\eea
is defined in terms of the set $L$, that includes all codewords 
\bea
\label{defL}
{\rm x}=(a_1,\dots, a_{g-\tilde{g}},b_1,\dots, b_{g-\tilde{g}})\in L\subset (\Z_N\times \Z_N)^{g-\tilde{g}}
\eea 
such that there exist $a_{g-\tilde{g}+1},\dots a_g$ satisfying 
\bea
\label{constr}
(a_1,\dots a_g,b_1,\dots b_{g-\tilde{g}},0\dots,0)\in \LL_\gamma. 
\eea
This definition makes it clear $L$ is closed under addition, i.e.~it is an additive code. 
The multiplicity $m({\rm c})$ is the number of different sets $a_{g-\tilde{g}+1},\dots a_g$ for any given ${\rm c}\in L$. 
We show in Appendix \ref{app:L} that $m({\rm c})=|M_0|$ is the same for all $\rm c$ and $L$ is a self-dual symplectic code of length $g'=g-\tilde{g}$.
With that we arrive at the desired result:  path integral of the 3d theory on any topology ending on $\Sigma$ of genus $g'$ is a CSS quantum stabilizer state defined by a classical slef-dual symplectic code $L$, 
\bea
\label{resultingstate}
{}^{\tilde{g}}\langle 0|U_\gamma|0\rangle^g=N^{-(g+\tilde{g})/2} |M_0| |L\rangle.
\eea

The relation between $|\LL\rangle$ and associated 3d topologies $X_3$, which could be many, can be understood geometrically using the representation of the wavefunction \eqref{5dtheta}. Focusing on $g=1$ for simplicity we readily see that $\LL$ parameterizes the cohomologies of $\Sigma=\partial X_3$ mod $N$. For example for $\LL_0=(*,0)$ the $a$-cycle of the torus is cohomologically trivial as which follows from the summation over all $a_i$ in  \eqref{5dtheta}. Similarly $\LL_1=(0,*)$ corresponds to a solid torus with the trivial $b$-cycle. When $N=p^2$ the code $\LL'=(p*,p*)$ corresponds to a non-handlebody topology with  torsion on both $a$ and $b$-cycles, see \cite{Dymarsky:2024frx} where this example is discussed in detail.  
Saying the same differently, a symplectic self-dual code $\LL$, through the Construction A \cite{conway2013sphere}, defines a symplectic self-dual lattice of one-cohomologies of $H_1(\Sigma,\Z)$, generalizing the construction of handlebodies explained in \cite{Maloney:2020nni}.

The statement and its derivation for the 5d theory \eqref{5d} is completely analogous. The vacuum state of $g=1$  theory -- the path integral on a 5d ``handlebody'' which is a direct sum of $n$ $B_3\times \SSS^2$ is 
\bea
|0\rangle_{5d}^n={1\over N^{n/2}}|\CC_0\rangle_{3d},\qquad (\alpha_1,\dots,\alpha_n,0,\dots,0)\in\CC_0,\qquad \alpha_i\in \Z_N. 
\eea
It is clearly a code state for  even self-dual code $\CC_0$. 
A mapping class group transformation, or more generally  any transformation from $h\in O(n,n,\Z_N)$, produces a state  
\bea
U_h|0\rangle^n ={1\over N^{n/2}}|\CC_h\rangle,
\eea
which is a code state associated with another even self-dual  code $\CC_h$,
\bea
c=(\alpha_1,\dots,\alpha_n,\beta_1,\dots,\beta_n)\in \CC_h,\quad {\rm iff}\quad  c=h(\alpha_1',\dots,\alpha'_n,0,\dots,0)\,\, {\rm mod}\,\, N.
\eea
The genus reduction of this state is given by 
\bea
\label{resultingstate5d}
{}^{\tilde{n}}\langle 0|U_h|0\rangle^n=N^{-(n+\tilde{n})/2} |M_0| |L\rangle,
\eea
where sets $L_i,M_i$ are defined similarly to the 3d case discussed in the Appendix \ref{app:L}. It is straightforward to see from the definition that $L$ is an even code. To show that it is self-dual, one can follow the same logic as in the 3d case, evaluating its size $|L|=N^{n-\tilde{n}}$.

The geometric relation between $|\CC\rangle$ and $X_5$ can be established using the wavefunction representation \eqref{3dtheta} in a way similar to the 3d case discussed above: the code $\CC$ parameterizes two-cohomologies of $\M_4=\partial X_5$. 

Using the isomorphism between 3d and 5d theories, we can formulate our findings above as follows.  Any state produced by a path integral of the 3d theory on any given 3d topology is (up to an overall coefficient) the state produce by a topological boundary condition in 5d theory. Similarly, any state produced by the 5d path integral on any 5d topology is the state in 3d theory produced by topological boundary conditions, again up to an overall normalization. It would be very tempting to give this observation a geometric interpretation in terms of the 7d theory \eqref{7d}.

\section{Holographic correspondence}
\label{sec:holography}

The AdS/CFT correspondence, also known as holographic duality, is an equivalence between a (non-gravitational) field theory on living on a  $d$-dimensional ``boundary'' manifold $\M$, and a (gravitational) field theory in $d+1$ dimensions  on  a ``bulk'' manifold $X_{\rm bulk}$  that ends on $\partial X_{\rm bulk}=\M$. Given that the bulk theory is gravitational, the bulk path integral includes a sum over bulk manifolds of all possible topologies satisfying $\partial X_{\rm bulk}=\M$ \cite{Witten:1998zw}.
The statement of holographic duality is often formulated as 
\bea
\label{holography}
Z_{\rm CFT}[J]=Z_{\rm bulk}[J],
\eea
where both bulk and boundary  partition functions depend on $J$. The latter has different interpretations on the two sides of the duality. On the CFT side, $J$ represents the classical external sources,  while on the bulk side $J$ specifies boundary conditions at $\partial X_{\rm bulk}=\M$ in the bulk path integral. We emphasize that on both sides $J$ represents classical data ``living'' on $\M$, which in turn specifies the coupling constants of the $d$-dimensional theory. 

An older idea, originated in \cite{Witten:1988hf} and known as the TQFT/(R)CFT correspondence, establishes a similar yet distinct relation between bulk path integrals in a 3d TQFT and conformal blocks in a dual 2d CFT. Unlike the AdS/CFT correspondence, this relation, commonly known as the bulk-boundary correspondence, does not involve a sum over bulk geometries.   To illustrate it, consider a $(2+1)$-dimensional topologically ordered system on a spatial disc $D$ described by a TQFT $\mathcal{T}$. Depending on the boundary conditions at $\partial D$, such a system may exhibit massless edge modes. These modes can be  described either by a $(1+1)$-dimensional theory on $\partial D\times \R$ or by a $(2+1)$-dimensional  theory $\mathcal{T}$ on the space $X_{\rm bulk}= D\times \R$ with boundary $\partial D\times \R$ \cite{Witten:1988hf,Elitzur:1989nr}. 

\begin{figure}
    \centering
    \includegraphics[width=0.8\linewidth]{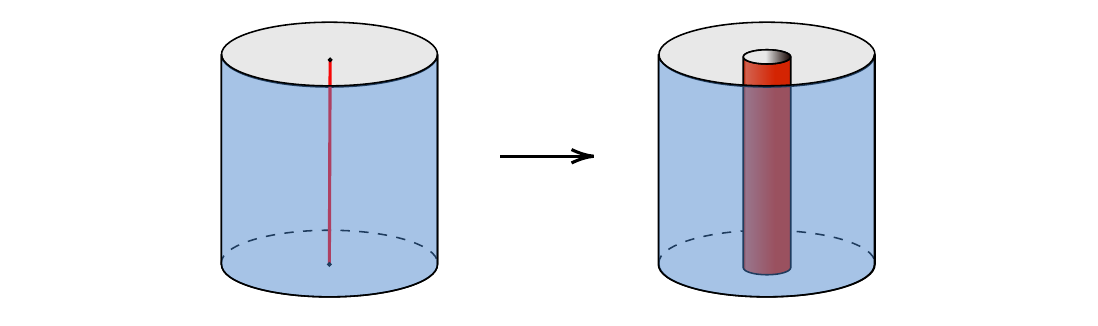}
    \caption{Left image:  a $2+1$ topological system with a defect. A particular combination of defects gives rise to topological boundary conditions at the defect worldline and a CFT at the cylinder boundary. Right image: the same, after the defect worldline is fattened into a cylindrical shell (shown in red). The result is the sandwich construction: topological boundary condition at the boundary shown in red gives rise to a CFT at the {\it other} boundary. }
    \label{fig:fattened_worldline}
\end{figure}

This scenario is of course very similar to the holographic duality described above, and it is frequently called holography in the condensed matter literature, but there is a crucial distinction that we would like to emphasize. The path integral in $2+1$ dimensions evaluates a particular conformal block of the $(1+1)$-d theory, not the entire modular-invariant partition function. In other words certain sectors of the $(1+1)$-d theory are missing. It is well known that to generate excitations in those sectors on the boundary, the $(2+1)$-d theory on the disc should be amended to include corresponding defects. Certain combinations of defects, such that the bulk fields have topological (gapped) boundary conditions at the defect worldline, will give rise to a full, modular-invariant theory at the boundary $\partial D$ (without any light dynamical modes living on the defect itself). Since the bulk theory $\mathcal{T}$ is topological, the defect worldline can be fattened into $\SSS^1\times \R$ as shown in Fig.~\ref{fig:fattened_worldline}, so that the disc $D$ becomes an annulus. As a result, we obtain the description of the boundary CFT on $\partial D\times \R$ in terms of the so-called sandwich construction \cite{Ji_Wang:2019jhk_symtft,Gaiotto:2020iye,Apruzzi:2021nmk_symtft,Freed:2022qnc_symtft}: the topological theory  $\mathcal T$ on the annular cylinder $(\partial D\times \R)\times [0,1]$ with topological boundary conditions at one of the boundaries gives rise to a CFT at the other. Often a topological theory admits several different possible choices of topological boundary conditions $\cal C$ (if it admits any). Choosing any one of them at one boundary (while the boundary conditions at the other boundary should be chosen to admit massless modes) and evaluating the path integral on the cylinder would yield the partition function of the CFT specified by $\cal C$,
 \bea
 Z_{\cal C}=Z_{\rm bulk}.
 \eea
This is of course similar to holography. But the crucial difference between  holography and the sandwich construction is that, in the latter case, the details of the boundary theory are specified by the boundary conditions of the bulk fields at the {\it other} boundary, not the one where the boundary theory lives.  This difference is both geometric and conceptual.  Given the topological nature of $\mathcal T$, the ``width'' of the sandwich can be made arbitrarily small resulting in a $d$-dimensional theory \cite{Kapustin:2014gua}. In the case of holography the additional direction is dynamical and the bulk theory is $(d+1)$-dimensional. Another important differences is that in holography bulk theory is gravitational and hence includes a sum over all topologies of $Z_{\rm bulk}$.

There is a particular scenario in which the distinction between the bulk-boundary correspondence and holography disappears --- when the theory $\mathcal T$ is topologically trivial and non-anomalous. In this case its path integral on $D\times \R$ evaluates the 
modular-invariant 2d CFT partition function. This was first pointed out in \cite{gukov2005chern}, which formulated a holographic description of a 2d Narain (compact scalar) CFT in terms of a trivial 3d BF-type Abelian Chern-Simons theory of level $k=1$. In this paper we revisit this idea and provide a holographic description for 4d Maxwell theory in terms of a trivial  5d BF-type Abelian TQFT of level $N=1$. 

\begin{figure}
    \centering
    \includegraphics[width=1\linewidth]{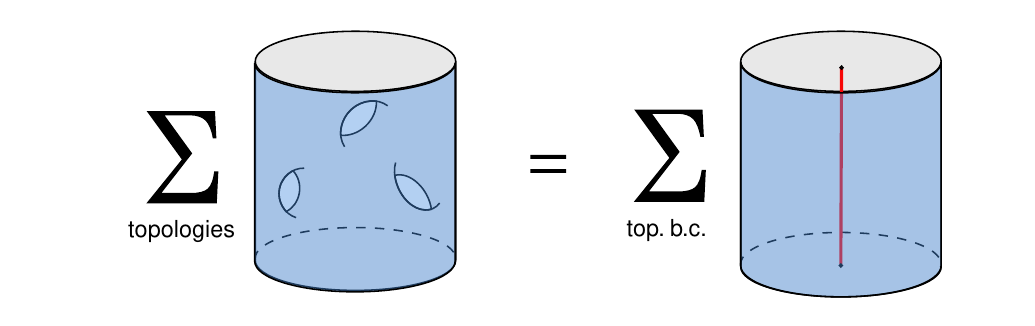}
    \caption{A schematic illustration of the sum over 3d topologies being equal to the sum over topological boundary conditions. The sum includes all 3d topologies, smooth topologies (handlebodies) as well as singular ones, obtained via genus reduction.  Possible weights on both sides of the equality are omitted for visual simplicity.}
    \label{fig:sumsum}
\end{figure}

Yet, the main approach we take in this paper is different. Inspired by \cite{Castro:2011zq, Maloney:2020nni, Afkhami-Jeddi:2020ezh}  and following \cite{Barbar:2023ncl,Dymarsky:2024frx} we promote $\mathcal T$ to a ``gravitational'' theory -- TQFT gravity -- by summing over all possible topologies of $X_{\rm bulk}$ with fixed $\partial X_{\rm bulk}$ when evaluating the path integral. Thus, in the case discussed above the path integral will be evaluated and summed over all 3d topologies ending on $\partial D\times \R$. 
We emphasize that, as in the case of conventional holography, the bulk has only one boundary $\partial D$, where the CFT lives, and where boundary conditions for the bulk fields are specified. It has been shown in \cite{Dymarsky:2024frx} that summing over all possible topologies of $X_{\rm bulk}$ is mathematically the same as inserting a linear combination with {\it positive} coefficients of all possible topological boundary conditions at the internal boundary of the sandwich construction, as we schematically illustrate in Fig.~\ref{fig:sumsum}. After summing over topologies the bulk path integral will yield an ensemble-averaged CFT partition function 
\bea
\label{ensembleholography}
\langle Z_{\rm CFT}\rangle\equiv \sum_\CC \alpha_\CC Z_{\CC}=  Z_{\rm TQFT\, gravity},\qquad \alpha_\CC\geq 0.
\eea
This identity represents a holographic duality between TQFT gravity in the bulk and an ensemble of CFTs on the boundary. 

\subsection{Preliminaries of 4d Maxwell theory}


We first recall the preliminaries of 4d Maxwell theory.  On a spin four-manifold $\M_4$ Maxwell is characterized just by the coupling constant $\tau={4\pi i\over g^2}+{\theta\over 2\pi}$,\footnote{Maxwell theory on non-spin manifolds is significantly more complicated as has been discussed in \cite{Metlitski:2015yqa,Hsin:2019fhf,Ang:2019txy,Brennan:2022tyl,Davighi:2023luh,Kan:2024fuu}.}
\bea
S=\int_{\M_4} {1\over g^2} F\wedge \star F +{i\theta \over 8\pi^2} F\wedge F.
\eea
Theories with $\tau$ and $\tau'$ related by an $SL(2,\Z)$ transformation are S-dual to each other, i.e.~they are physically equivalent, although due to an anomaly the partition functions 
$Z_\tau[\M_4]$ and $Z_{\tau'}[\M_4]$ may differ by a phase \cite{witten1995s,Etesi:2010sb,Seiberg:2018ntt,Hsieh:2019iba}. When the manifold has vanishing signature $\sigma=0$ the anomaly cancels. The partition function was evaluated in \cite{verlinde1995global,witten1995s} to be\footnote{There is an ambiguity in the power of $\tau_2$ inside $\theta$  associated with the freedom to add a local term to the action \cite{witten1995s}. We fix it by requiring $Z_\tau$ to be invariant under S-transformation $\tau\rightarrow -1/\tau$.} 
\bea
\label{Z4d}
Z_{\tau}[\M_4]={\theta(\tau,E,\xi,\bar\xi)\over  \Phi_{5d}},\quad \Phi_{5d}={({\rm det'}\Delta_1)^{1/2}\over {\rm det'}\Delta_0}.
\eea
Here $\theta$ is a Siegel-Narain theta function defined in terms of the ``Narain data'' $E=G+B$, specified by the four-manifold $\M_4$, see Appendix \ref{sec:4dgeometry} for details, 
\bea
\label{MaxwellZ}
\theta(\tau,E,\xi,\bar\xi)&=&\tau_2^{n/2}\sum e^{i\pi \tau p_L^2-i\pi\bar\tau p_R^2+2\pi i(p_L\xi-p_R\bar\xi)+{\pi\over 2\tau_2}(\xi^2+\bar \xi^2)},\\
&&{1\over \sqrt{2}}
\left(\begin{array}{c}
p_L^I+p_R^I\\
p_L^I-p_R^I
\end{array}
\right)= \mathcal{O}\, \vec{n},\qquad \vec{n}\in \Z^{2n},
\eea
and $\mathcal O$ is given by \eqref{Omatrix}. The same theta-function after Poisson resummation can be written as follows,
\bea
\label{MaxwellZtilde}
\tilde{\theta}(\tau,E,z,\bar z)&=&{\rm det}(G)^{1/2}\sum
e^{-{\pi}|{\rm v}|^2-2\pi i n^T Bm +\sqrt{2}{\pi}(\bar z^T  {\rm v}-z^T {\rm v}^*)+{\pi}\bar z^T z },\\
&&{\rm v}=G^{1/2}(n+\tau m)/\sqrt{\tau_2},\quad n,m\in \Z^n,
\eea
where 
\bea
\label{zdef}
\theta(\tau,E,\xi,\bar\xi)=\tilde{\theta}(\tau,E,z,\bar z),\qquad \xi=z\sqrt{\tau_2},\, \bar\xi=\bar z \sqrt{\tau_2}.
\eea
The classical sources $G^{1/2}z$ and $G^{1/2}\bar z$ couple to $\U (1)\times \U (1)$-charges of Maxwell theory: fluxes of  $(F^D-\bar \tau F)/\sqrt{\tau_2}$ and its conjugate evaluated through half of the two-cycles of $\M_4$. Here $F^D=\tau_2 \star F+\tau_1 F$ and $F$ form a doublet under $SL(2,\Z)$.  The lattice-independent factor $e^{\pi \bar z^T z}$ in \eqref{MaxwellZtilde} arises because we are evaluating field theory path integral, not the partition function; see \cite{Aharony:2023zit,Kraus_2007} for a detailed discussion of the difference between the two in the 2d case. 

Appearance of the Siegel-Narain theta function in \eqref{Z4d}, which is  up to the factor of $\Phi$ the same as  the partition function of 2d Narain CFT, can be understood directly in field theory, by considering six-dimensional theory of the self-dual two-form on $\Sigma\times \M_4$ \cite{verlinde1995global,witten2004conformal}. 

A generalization of the above discussion to Maxwell theory with gauge group $\U (1)^g$ is straightforward. In this case the coupling is a $g\times g$ ``modular parameter'' $\Omega$, and the partition function on a spin four-manifold $\M_4$ is given by 
\bea
Z_{\Omega}[\M_4]={\theta(\Omega, E,\xi,\bar\xi)\over \Phi_{5d}}={\tilde\theta(\Omega, E,z,\bar z)\over \Phi_{5d}},\quad \xi=z\,\Omega_2^{1/2},\, \bar\xi=\bar z\,\Omega_2^{1/2},
\eea
where $\theta,\tilde\theta$ are given by \eqref{3dtheta},\eqref{5dtheta} with $N=1$ and vanishing $c_i,{\rm c}_i$. 
We note that $\theta(\Omega,E,\xi,\bar \xi)=\tilde \theta(\Omega,E,z,\bar z)$ is invariant under modular transformations \eqref{modulartransformation}, reflecting S-duality of 4d Maxwell theory, as well as under orthogonal transformations \eqref{orthogonaltransformation}, which are the mapping class group transformations of the 4d manifold $\M_4$.  Under modular transformations variables $z,\bar z$ change by a unitary matrix $U\in \U (g)= O(2n,\R)\cap Sp(2n,\R)$, 
\bea
z\rightarrow z_\gamma=z\, U,\quad \bar z\rightarrow \bar z_\gamma=\bar z\, U^\dagger,
\eea
defined by  polar decomposition of $\Omega_2^{1/2}(C\Omega+D)^{-1}$. Alternatively it can be defined by $S=H{\mathcal S}(\Omega_\gamma)\, \gamma\,{\mathcal S}^{-1}(\Omega)H^{-1}$, in full analogy with the matrices $O_L,O_R$ enacting orthogonal transformations \eqref{orthogonaltransformation}, $z\rightarrow z_h=O_L\, z$, $\bar z\rightarrow \bar z_h=O_R\, \bar z$.

The representation \eqref{MaxwellZtilde} and its generalization to higher $g$, given by \eqref{5dtheta} with $N=1$, provides an alternative way to think about the Siegel-Narain theta function in $\R^{n,n}$. Essentially it is a lattice theta function for a self-dual {\it symplectic} lattice $\Z^{2g}\in \R^{2g}$ equipped with the symplectic inner product \eqref{J} and conventional Euclidean metric. All such lattices are parameterized by modular parameters $\Omega$ subject to $\Sp(2g,\Z)$ identifications \cite{symplecticlattices}, in terms of which the lattice-generating matrix can be chosen as follows:
\bea
\label{symplecticlattice}
\vec{\rm v}={\mathcal S}\left(\begin{array}{c}
\vec{n}\\ \vec{m}\end{array}\right),\quad \vec{n},\vec{m}\in \Z^g,\quad 
{\cal S}=\Omega_2^{-1/2}\left(\begin{array}{cc}
\mathbb{1} &  \Omega_1\\
0 & \Omega_2\end{array}\right).
\eea
In what follows we will refer to the space of modular parameters (symplectic lattices) 
\bea
{\cal M}_g=Sp(2g,\Z)\backslash Sp(2g,\R)/\U (g,\R)
\eea
as the $\Sigma$-moduli space. This is the space of couplings of 4d $\U (1)^g$ gauge theories. 

Similarly to Narain lattice, which is a lattice of $\U (1)^n\times \U (1)^n$ charges of states in 2d theory, symplectic lattice above is the lattice of 
$\U (1)^g\times \U (1)^g$ charges of 4d Maxwell theory \cite{Kapustin:2005py}.

\subsection{$N=1$: 5d holographic dual to Maxwell theory}
We now proceed with the case of $N=g=1$ 5d theory \eqref{5d}, which is trivial in the TQFT sense: it has no non-trivial surface operators (topological defects) and its Hilbert space on any $\M_4$ is one-dimensional. Its path integral on any five-dimensional $X_{\rm bulk}$ with boundary $\partial X_{\rm bulk}=\M_4$ is   (up to an overall normalization) the same, and is given by \eqref{3dblock} with $c_1=0$ or \eqref{5dblock} with ${\rm c}_i=0$. It is straightforward to see that this is the same as the partition function of a 4d Maxwell theory on $\M_4$,
\bea
\label{HC4d}
Z_{\tau}[\M_4]=\Psi_0(\tau,E,\xi,\bar\xi)={\sf \Psi}_{0\dots 0}(\tau,E,\xi,\bar\xi)={\tilde\theta(\tau,E,z,\bar z)\over \Phi_{5d}},
\eea
where $\Psi_0$ is as in \eqref{3dblock} and ${\sf \Psi}_{0\dots 0}$ is defined in \eqref{5dblock}.
This relation, as well as a similar statement about 3d Chern-Simons theory and the 2d free boson, was already mentioned in \cite{verlinde1995global}, two years before the holographic correspondence was introduced. To our knowledge the statement about the TQFT in the bulk being holographically dual to the boundary CFT appears for the first time in \cite{Maldacena:2001ss}, without emphasizing that the TQFT should be trivial. The complete statement, that the trivial $AB$ theory \eqref{AB} with $N=1$ is dual to a compact scalar (an arbitrary Narain theory for $n>1$), was put forward in \cite{gukov2005chern}, and more recently revisited in \cite{Ashwinkumar:2021kav,benini2023factorization, Aharony:2023zit}. Holography for free field theories dual to trivial TQFTs in higher dimensions, including 4d Maxwell theory, was also discussed recently in \cite{Antinucci:2024bcm}. 

We have evaluated the RHS of \eqref{HC4d} explicitly in  section \ref{sec:holquantization} and matched it to the known field theory answer above. As we noted at the beginning of this section, the path integral in the bulk should in principle include a sum over all possible 5d bulk topologies $X_{\rm bulk}$ ending on $\M_4$, but in the case of a trivial topological theory that sum would merely introduce an overall coefficient, which is renormalized to ensure that the vacuum is unique, 
\bea
Z_{\rm bulk}(\tau,\xi,\bar\xi)=\Psi_0(\tau,E,\xi,\bar\xi)\equiv {\sf \Psi}_{0\dots 0}(\tau,E,\xi,\bar\xi).
\eea
We thus obtain a conventional statement of holography \eqref{holography} for 4d Maxwell theory, which takes the form 
\bea
Z_{\tau}[\M_4]=Z_{\rm bulk}(\tau,\xi,\bar\xi).
\eea
The generalization of this relation to arbitrary gauge group $\U (1)^g$ is straightforward.  

Since the $N=1$ theory is topologically trivial, the bulk path integral on $X_{\rm bulk}$ factorizes when $\M_4=\partial X_{\rm bulk}$  is a union of several disconnected manifolds. This is no longer true for $N>1$.

\subsection{$N>1$: holographic dual to an ensemble}
\label{sec:ensemblehol}

Before we formulate the duality for $N>1$, it is helpful to revisit the 3d/2d case, in which a 3d TQFT summed over all three-dimensional topologies ending on $\partial X_{\rm bulk}=\Sigma$ evaluates the averaged partition function of an ensemble of 2d CFTs. This proposal in its current form was put forward in \cite{Barbar:2023ncl} and proved for a general 3d TQFT with a finite number of anyons in \cite{Dymarsky:2024frx}, which gave a precise definition of the sum over topologies and determined the weights of the boundary ensemble. An Abelian example, relevant to what follows, was discussed in detail for a torus boundary manifold   in \cite{Aharony:2023zit} and for general a general $\Sigma$ in \cite{Dymarsky:2025agh}. 

We start with a 3d TQFT in the sandwich construction, as reviewed above. Each topological boundary condition $\cal C$ in this theory gives rise to a partition function of a 2d CFT on $\Sigma$ 
\bea
\label{ZBC}
Z_\CC=\langle {\cal B}|\CC\rangle, 
\eea
where the appropriate conformal boundary conditions on $\Sigma$ are encoded in $\langle {\cal B}|$. For an Abelian theory, $\CC$ can be associated with an even self-dual code and $|\CC\rangle$ is given by \eqref{Cstate}. For the $AB$ theory \eqref{AB} we explicitly define  the state  $\langle {\cal B}|=\langle \Omega,E,\xi,\bar \xi|$ such that 
\bea
Z_\CC(\tau,\xi,\bar\xi)=\langle \Omega,E,\xi,\bar \xi|\CC\rangle={\theta(\Omega,E_\CC,\xi,\bar\xi)\over \Phi_{3d}}  
\eea
is the genus-$g$ path integral of a Narain theory specified by the point in the Narain moduli space $E_\CC$. The relation between a code $\CC$ and $E_\CC$ is as follows. The code $\CC$ defines a Narain lattice $\Lambda_\CC$ via the so-called Construction A, see \cite{Angelinos:2022umf,Aharony:2023zit,Barbar:2023ncl} for details, which in turn defines  $E_\CC=G_\CC+B_\CC$.

We note, however, that the statement \eqref{ZBC} is more general and holds in any TQFT  that admits topological and conformal boundary conditions. The boundary ensemble is an ensemble of CFTs corresponding to all possible topological boundary conditions $\CC$, 
\bea
\label{boundaryensemble}
\langle Z\rangle=\sum_\CC \alpha_\CC Z_{\CC},
\eea
with positive weights $\alpha_\CC$ reflecting the size of the symmetry group of each theory \cite{Barbar}. In the Abelian case all the $\alpha_\CC$ are equal and can be normalized to $\alpha_\CC=\NC^{-1}$ where $\NC$ is the total number of topological boundary conditions.  

With an appropriate overall normalization \eqref{boundaryensemble} is equal to the path integral of the original TQFT summed over all 3d manifolds $X_{\rm bulk}$ ending on $\Sigma$. This sum is defined in terms of the mapping class group of a $\Sigma'$ of infinite genus, related to $\Sigma$ by genus reduction $\Sigma'\rightarrow \Sigma$ \cite{Dymarsky:2024frx}. In general, the resulting sum will include singular topologies. 

As discussed in section \ref{sec:topologyandcodes},
in the case of the 3d Abelian theory \eqref{AB} a state on $\Sigma$, evaluated by a path integral on a given $X_{\rm bulk}$ ending on $\Sigma=\partial X_{\rm bulk}$, is a stabilizer state specified by a self-dual symplectic  code $\cal L$. The resulting statement of ensemble holography can then be formulated as follows,
\bea
\label{hbulk}
{1\over \NC}\sum_{\cal C}|\CC\rangle=\sum_{\LL} \beta_\LL |\LL\rangle,
\eea
where the coefficients $\beta_\LL$, which result from the sum over topologies, require evaluation. Comparing with \eqref{boundaryensemble} we note that $\langle {\cal  B}|$ has been stripped from both sides of the identity, and thus the resulting mathematical statement is about states of the TQFT on $\Sigma$.

In the specific case of an $AB$ theory \eqref{AB} with $N=\prod_k p_k$ a square-free product of distinct primes $p_k$, the identity \eqref{hbulk} simplifies to 
\bea
\label{Nprime}
{1\over \NC}\sum_{\cal C}|\CC\rangle= {A\over \NL}\sum_\LL  |\LL\rangle.
\eea
This identity follows from the uniqueness of the state invariant under both symplectic \eqref{simplgroupaction} and orthogonal \eqref{orthgroupaction} groups, and can be understood in terms of Howe duality \cite{Dymarsky:2025agh}. The coefficient $A$ can be evaluated to be
\bea
A={1 \over \NC } \prod_k p_k^{-ng/2}\prod_{i=0}^{n-1} (p_k^i+p_k^g).
\eea 
and $\NL$ is the number of symplectic self-dual codes $\LL$ of length $2g$, evaluated in Appendix \ref{app:p2counting}.

When $N$ is not square-free there is generally more than one invariant vector and the expression is more involved. Focusing on the case $N=p^2$ for prime $p$ and taking $g=1$, we find that there are two invariant vectors, one for each orbit $S_a$ of $\Sp(2g,Z_N)$ acting on symplectic codes,   and the identity \eqref{hbulk} takes the form
\bea
\label{3dholographyp2}
{1\over \NC}\sum_{\cal C}|\CC\rangle=A_{p^2} \sum_{\LL\in S_1} |\LL\rangle + B_{p^2} \sum_{\LL\in S_0} |\LL\rangle.
\eea
The coefficients $A_{p^2},B_{p^2}$ are evaluated in Appendix \ref{matchingRHS}. 

Now returning to 5d, the ensemble of boundary theories is defined by the set of all topological boundary conditions $\LL$ of the 5d theory. As we discussed in section \ref{sec:SymTFTpartI}, in the case of interest this is the set of self-dual symplectic codes and the analog of \eqref{ZBC} takes the form 
\bea
\label{ZL}
Z_\LL=\langle {\cal B}_{4d}|\LL\rangle.
\eea
Here $\langle {\cal B}_{4d}|=\langle \Omega,E,\xi=z\,\Omega_2^{1/2},\bar \xi=\bar z\,\Omega_2^{1/2}|$ and 
\bea
Z_\LL\equiv Z_{\Omega_\LL}[\M_4,z,\bar z]={\tilde\theta(\Omega_\LL,E,z,\bar z)\over \Phi_{5d}}
\eea
is the path integral of a Maxwell theory on $\M_4$ characterized by $E$, with coupling constant
$\Omega_\LL$ and external sources $z,\bar z$. The relation between a self-dual symplectic code $\LL$ and $\Omega_\LL$ is as follows. The code defines a self-dual symplectic lattice in $\R^{2g}$ via Construction A \cite{conway2013sphere}
\bea
\vec{\rm v}={\cal S}(\Omega)\left(\begin{array}{c}
N\,\vec{n}+\vec{a}\\
N\,\vec{m}+\vec{b}\end{array}\right)/\sqrt{N},\quad {\vec n},\vec{m}\in \Z^g,\quad (\vec{a},\vec{b})\in \LL,
\eea
and $\Omega_\LL$ is the associated modular parameter of this lattice. 

To derive a 5d statement of holography analogous to \eqref{hbulk}, we follow the argument of \cite{Dymarsky:2024frx} and extend it where necessary. 
As shown in \cite{Nebe} in great generality, code states $|\LL\rangle$, interpreted as ``genus $n$'' full enumerator polynomials of self-dual classical codes $\LL$, span the space of $\Or (n,n,\Z_N)$-invariant states in $\HH$. It follows from \eqref{Lstate} that for a given $\M_4$ of zero signature and ``genus'' $n=b_2^+=b_2^-$, the overlap between two code states is given by 
\bea
\langle \LL|\LL'\rangle=|\LL\cap\LL'|^n, 
\eea
where $1\leq |\LL\cap\LL'|\leq N^g$ is the number of codewords present in both $\LL$ and $\LL'$. Provided the codes are distinct, this number is smaller than $|\LL|=|\LL'|=N^g$ and therefore for $n\rightarrow \infty$ all code states $|\LL\rangle$  become orthogonal. 
Therefore, in this limit,
\bea
\rho={1\over N^{gn} \NL} \sum_\LL {|\LL\rangle \langle \LL|} \in \HH^* \otimes \HH
\eea
is a projector on the $\Or (n,n,\Z_N)$-invariant subspace of $\HH$. The same projector can be written as the group average
\bea
\rho={1\over |O(n,n,\Z_N)|}\sum_{h\in O(n,n,\Z_N)} U_h. 
\eea
From here we find the identity valid in the large $b$ limit,
\bea
\label{state}
{1\over{N}_\LL} \sum_{\LL} |\LL\rangle ={ N^{gn} \over |O(n,n,\Z_N)|} \sum_{h\in O(n,n,\Z_N)} U_h|0\rangle_{5d},\qquad n\rightarrow \infty. 
\eea
When $N$ is prime, the map from $\Gamma_0(N)\backslash O(n,n,Z)$ to $\Gamma_0\backslash O(n,n,Z_N)$ is surjective 
in full analogy with the 3d case (where the $\Gamma_0(N)\backslash Sp(2g,Z)$ to 
$\Gamma_0\backslash Sp(2g,Z_N)$ is surjective for any $N$), and
this sum be interpreted as follows. The RHS is a sum over simple 5d geometries that generalize 3d handlebodies. The vacuum state $|0\rangle_{5d}$ is the TQFT path integral  on $X_5$, which is a connected sum of $n$ copies of $B^3 \times \SSS^2$, while the sum on $h$ runs over mapping class group transformations that generate all other ``handlebodies.'' In the case of general $N$,
$U_h$ may not have an interpretation as a mapping class transformation, but the resulting state $U_h|0\rangle_{5d}$ is a code state for an orthogonal even self-dual code $\CC_h=h\,\CC_0$ and hence is a path integral on a possibly singular 5d topology with a particular lattice of two-cohomologies.

To obtain a version with reduced ``genus" $n'$, the boundary $\M_4=\#^n (\SSS^2 \times \SSS^2)$ of a given bulk geometry can be attached to a cobordism that degenerates a nonintersecting subset of $n-n'$ boundary 2-cycles to zero, leaving a boundary  isomorphic to 
$\#^{n -\tilde n}(\SSS^2 \times \SSS^2)$.  
This procedure is the analog of genus reduction in 2d. Algebraically, it is represented by taking a scalar product of \eqref{state} with ${}^{n-n'}_{5d}\langle 0|\otimes  \langle {\cal B}_{4d}|$,
\bea
\label{holographic5d}
{1\over \NL } \sum_{\LL} Z_\LL =\lim_{n'\rightarrow \infty} { N^{gn'} \over |G_O|} \sum_{h\in G_O} {}^{n'-n}_{~~~5d}\langle 0| \langle {\cal B}_{4d}|U_h|0\rangle^{n'}_{5d},\quad G_O=\Gamma_0\backslash O(n',n',\Z_N).
\eea
Now the LHS is the desired ensemble-averaged CFT partition function on $\M_4$ while the RHS represents a sum over various 5d topologies ending on $\M_4$. This identity is the statement of holographic correspondence between the ensemble of 4d Maxwell theories on $\M_4$ and the dual 5d ``TQFT gravity'' -- the 5d theory \eqref{5d} summed over 5d topologies ending on $\M_4$.

The Heegaard splitting  theorem \cite{hempel20223} guarantees that the analog of the sum on the RHS in 3d includes all possible topologies with the given boundary. 
The five-dimensional case is more nuanced \cite{kim2025heegaard}. We proceed assuming the sum on the RHS of \eqref{holographic5d} includes all possible classes of the 5d topologies ending on $\M_4$ that can be distinguished by the Abelian theory \eqref{5d}. 

All terms appearing on the RHS of \eqref{holographic5d} are states associated with classical even self-dual codes $\CC$. This allows us to evaluate both sides explicitly by matching to the LHS for various $N$. A more general framework to perform such calculations was recently put forward in \cite{Nick}. 
For square-free $N$, we recover \eqref{Nprime} which can be written in a form making the 2d/4d duality manifest, 
\bea
\label{5dA}
&&{1\over \NL}  \sum_\LL  |\LL\rangle={A'\over \NC}\sum_{\cal C}|\CC\rangle,\\
&& A'={1 \over \NL }
\prod_k p_k^{-ng/2}\prod_{i=1}^g (p_k^i+p_k^n),\qquad AA'=1.
\eea
Focusing on the case of a single U$(1)$ gauge field ($g=1$) and prime $N=p$, this can be written as 
\bea
\label{holsum5d}
&&\langle Z_{4d\,\rm Maxwell}\rangle={1\over \NL} \sum_\LL Z_{\tau_\LL}[\M_4,z,\bar z]=\tilde{A}\sum_{h\in \Gamma_0\backslash O(n,n,\Z_N)} {\sf \Psi}_{0\dots 0}(\tau, E_h\, \xi_h, \bar\xi_h),\\
&& \xi=\sqrt{\tau_2}z,\, \bar\xi=\sqrt{\tau_2}\bar z,\quad  \tilde{A}={p+p^n\over \NL\NC},\\
&& \NL=(p+1),\qquad \NC=|O(n,n,\Z_N)|/|\Gamma_0|=\prod_{i=0}^{n-1}(p^i+1),
\eea
where the sum in the LHS of \eqref{holsum5d} goes over $\NL=p+1$ possible values of the coupling constant
\bea
\label{tauL}
\tau_{0}=p\tau,\qquad  \tau_{{r+1}}={\tau+ r\over p},\,\quad r=0\dots p-1.
\eea

In disguise, this is the same mathematical identity as the one in 3d discussed in \cite{Aharony:2023zit}, but with the interpretation of the LHS and RHS interchanged. The sum in the LHS of \eqref{holsum5d} is over an ensemble of Maxwell theories in 5d, while in 3d it was a sum over 3d topologies. Similarly, the sum over orthogonal group in the RHS is the sum over 5d ``handlebodies,'' while in 3d this was a sum over points in the Narain moduli space. We make the relation to 3d manifest in Section \ref{sec:largeN} below. 


We stress that  the holographic identity in 5d (\ref{5dA}) being mathematically the same as  the 3d identity (\ref{Nprime}) for any $g,n$ is a coincidence specific for square-free $N$. For general $N$ the identity takes the form  
\bea
\label{holid5d}
{1\over \NL}  \sum_\LL  |\LL\rangle={1\over \NC}\sum_{k} \alpha_k \sum_{{\cal C}\in S_k}|\CC\rangle,
\eea
where coefficients $\alpha_k$ are $g,n,N$-dependent and $S_k$ are distinct orbits of $\Or (n,n,\Z)$. It is different from its counterpart in 3d \eqref{hbulk}. 
In general for $g>n$ coefficients $\alpha_k$  can not be fixed uniquely because of possible degeneracy of the states in the RHS of \eqref{holid5d}. 
For $N=p^2$, $g=1$ and arbitrary $n$ we find coefficients $A_{p^2}',B_{p^2}'$ in the Appendix \ref{matchingRHS},
\bea
{1\over \NL}  \sum_\LL  |\LL\rangle=A'_{p^2}  \sum_{{\cal C}\in S_n}|\CC\rangle+B'_{p^2}\sum_{{\cal C}\in S_0}|\CC\rangle.
\eea
where we made the choice to keep all $\alpha_k=0$ except for $k=0,n$.

\subsection{Large-$N$ limit}
\label{sec:largeN}
For square-free $N$, when the orthogonal and symplectic groups act transitively on the corresponding codes, the holographic identity \eqref{Nprime} and \eqref{5dA} can be rewritten as an equality between linear combinations of theta functions
(below we often omit $\bar z$ for notational simplicity)
\bea
{1\over \NC}\sum_\CC \tilde\theta(\Omega,E_\CC,z)={A\over \NL}\sum_\LL \tilde\theta(\Omega_\LL,E,z).
\eea
On the LHS the theta function is averaged over a discrete set of points in the Narain moduli space; on the RHS the average is over points in the $\Omega$ moduli space ${\cal M}_g$. As $N$ increases, the number of points in each set also increases (with a possible exception of $\NC$ for $n=1$). As $N\rightarrow \infty$, the points densely and uniformly populate the corresponding moduli spaces. In the limit, depending on whether $n$ is larger or smaller than $g+1$, either the average over ${\cal M}_g$ or over Narain moduli space will diverge, with the leading behavior captured by real Eisenstein series of the symplectic or orthogonal groups correspondingly. This divergence will be compensated by $A$ that will also  diverge or go to zero. The resulting identity is a version of the Siegel–Weil formula, that made an appearance in the 3d case \cite{Afkhami-Jeddi:2020ezh,Maloney:2020nni} as well as in studies of string loop amplitudes \cite{Obers:1999um,Obers:1999es,Pioline:2001jn}. 

When $N$ is not square-free, the identities for the 3d and 5d theories take a slightly different form: points in the moduli space associated with {\it field theories} in 2d and 4d correspondingly are weighted equally, while points in the moduli space associated with topologies acquire nontrivial weights. As we discuss below, we find that the resulting identities in the $N\rightarrow \infty$ limit still take the same general form.

To derive the picture sketched above we start with square-free $N$ and  and evaluate the  scalar product of \eqref{5dA} with $\langle {\cal B}|$ (taking $g=1$ for simplicity),
\bea
\label{LHS}
{1\over \NL} \sum_\LL \langle {\cal B}|\LL\rangle={1\over \NL \Phi}\sum_\LL \tilde\theta(\tau_\LL,E,z).
\eea
 Taking $\Phi=\Phi_{5d}$ we  interpret this equation as an  average over an ensemble of Maxwell theories. Another representation, and interpretation, comes from the fact that for a square-free $N$ all codes $\LL$ belong to the same orbit under $SL(2,\Z)$. Combined with  the relation  
\bea
{\Psi}_{0\dots 0}(\Omega, E,\xi)=
\langle {\cal B}|0\rangle_{3d}={1\over N^{ng/2}}
\langle {\cal B}|\LL_0\rangle={1\over N^{ng/2}}{\theta(N\tau, E,\sqrt{N}\xi)\over \Phi}.
\eea
this gives 
\bea
\label{LHS2}
{1\over \NL} \sum_\LL\langle {\cal B}|\LL\rangle={N^{n/2}\over \NL}\sum_{\gamma \in G_S} { \Psi}_{0\dots 0}(\tau_\gamma, E,\xi_\gamma)={1\over \NL\, \Phi} \sum_{\gamma\in G_S} \theta(N \tau_\gamma, E, \sqrt{N}\xi_\gamma),\\ \nonumber
\NL=|G_S|,\qquad G_S=\Gamma_0\backslash SL(2,\Z_N).
\eea
Here $\tau_\gamma,\xi_\gamma$ are the modular transformations of $\tau,\xi$ given by \eqref{modulartransformation}. Up to an overall normalization the sum in the middle of \eqref{LHS2} has the interpretation in 3d as the sum over handlebody geometries obtained by a modular transformation $\gamma$ from the handlebody $X_3$ used to define the basis \eqref{3d}.
To obtain \eqref{LHS} from the rightmost expression in \eqref{LHS2} we note that $\theta$ is modular invariant 
and for a square-free $N$ there are always $\gamma',\gamma\in SL(2,\Z)$ such that
\bea
\label{gammagammap}
\gamma' \left(\begin{array}{cc}
N & 0\\
0 & 1\end{array}\right)\gamma=\left(\begin{array}{cc}
p & r\\
0 & q\end{array}\right),
\eea
with $pq=N$ and $0\leq r\leq q-1$. From here follows the expression for $\tau_\LL=(N\tau_\gamma)_{\gamma'}$,
\bea
\tau_\LL={p\tau+r\over q},\quad pq=N,\quad r=0\dots q-1,
\eea 
for all possible $p\, |\, N$.
For a prime $N$ this reduced to  \eqref{tauL}. A direct check shows that \eqref{gammagammap} also implies
\bea
(\sqrt{N}\xi_\gamma)_{\gamma'}=\sqrt{p\over q}\xi 
\eea
and therefore $z=\xi/\sqrt{{\rm Im\,} \tau}=(\sqrt{N}\xi_\gamma)_{\gamma'}/\sqrt{{\rm Im\,}\tau_\LL}$ remains the same for all terms in \eqref{LHS}.

The explicit values of $\tau_\LL$ given by \eqref{tauL} and the construction of the sum over $\gamma\in G_S$ above readily suggests that when the sources $z=\bar z=0$ vanish the sum can be written in terms of the  Hecke operator $T_N$ \cite{koblitz2012introduction} acting on the modular form of weight zero \cite{Aharony:2023zit}, 
\bea
\label{Hecke}
\sum_{\gamma\in SL(2,\Z)\backslash M_N}\tilde\theta(\gamma \tau, E,0)=\sum_\LL \theta(\tau_\LL,E,0)=N\, T_N\,   \theta(\tau,E,0).
\eea
Here $M_N$ is the space of  integer-valued $2\times 2$ matrices of determinant $N$. 

The generalization to arbitrary $g$ and $N$ is straightforward, 
\bea
\label{LHSg}
{\Phi \over \NL} \sum_\LL\langle {\cal B}|\LL\rangle={N^{ng/2}\over \NL}\sum_{\gamma \in G_S} { \Psi}_{0\dots 0}(\Omega_\gamma, E,\xi_\gamma)={1\over \NL} \sum_{\gamma\in G_S} \tilde\theta(N \Omega_\gamma, E, z_\gamma)=\\
{1\over \NL} \sum_{\LL} \tilde\theta(\Omega_\LL, E, z),\qquad 
\NL=|G_S|,\qquad G_S=\Gamma_0\backslash Sp(2g,\Z_N).
\eea
And again for vanishing sources $z=\bar z=0$ the sum can be represented in terms of a Hecke operator acting on $\tilde\theta(N\Omega,E,0)$.

When $n>g+1$, the sum $\sum_\gamma \Psi_{0\dots 0}(\Omega_\gamma, E,\xi_\gamma)$  converges in the $N\rightarrow \infty$ limit,
but the coefficient $N^{ng/2}/\NL$ diverges indicating 
 that average over $\cal M_g$ is singular. 
 For $N\gg 1$ and for fixed $\Omega$ only origin of the lattice contributes to $\theta(N\Omega,E,z)$, such that
\bea
\label{approxpsi0}
\Psi_{0\dots 0}(\Omega, E,\xi)\approx {\rm det}(\Omega_2)^{n/2}e^{{\pi\over 2}\Omega_2^{-1}(\xi^2+\bar\xi^2)}. 
\eea
Summing over $\gamma$ yields the generalization of the real Eisenstein series \cite{Datta:2021ftn},
\bea
\label{Eis}
\sum_{\gamma \in G_S} { \Psi}_{0\dots 0}(\Omega_\gamma, E,\xi_\gamma)={\rm det}(\Omega_2)^{n/2} E_n(\Omega,z),\qquad \xi=z\,\Omega_2^{1/2},\\
E_n(\Omega,z,\bar z)=\sum_{\gamma \in \Gamma_0\backslash Sp(2g,\Z)} { e^{{\pi\over 2} (z\, U^2 z^T +\bar z (U^\dagger)^2 \bar z^T)}\over |{\rm det}(C\Omega+D)|^n}.
\eea
We remind the reader that $\U (\gamma,\Omega)$ is defined through the polar decomposition of $\Omega_2^{1/2}(C\Omega+D)^{-1}$.
To carefully justify \eqref{Eis} one should take into account that the approximation \eqref{approxpsi0} is only valid if $\Omega$ is not too small, which could happen in the sum over $\gamma$ when matrices $C,D$ are sufficiently large. An argument that one can always choose a representative $\gamma \in \Gamma_0\backslash SL(2,\Z_N)$ with $|c|,|d|$ not exceeding $\sqrt{N}$ for a prime $N$ and thus completing the argument was given in \cite{Aharony:2023zit}. Generalization to higher $g$ is an open question.  

When $g>n-1$ the sum over $\gamma$ diverges, but the Hecke representation suggests that after normalizing by $\NL$ it is given by the average of $\tilde\theta$ over the fundamental domain of $\Omega$,
\bea
{1\over \NL} \sum_{\LL} \tilde\theta(\Omega_\LL, E, z,\bar z )=\langle \tilde\theta(\Omega,E,z\, U, \bar z\, U^\dagger)\rangle_{\Omega,U}\equiv \int {dU\over V_{U}} \int {d^{2g}\Omega \over  V_{g}({\rm det}\,\Omega_2)^2} \tilde\theta(\Omega,E,z U, \bar z U^\dagger).\nonumber
\eea
where 
$$
V_U=\prod_{k=1}^g \frac{2\pi^k}{\Gamma(k)}\qquad{\rm and}\qquad V_{g}=\prod_{k=1}^g \frac{2\Gamma(k)
\zeta(2k)}{\pi^k}
$$
are the Haar volumes of $\U(g)$ and ${\cal M}_g$, respectively \cite{siegel2014symplectic}.  
This is because the Hecke points, e.g.~\eqref{Hecke} for $g=1$ and $N\rightarrow \infty$,  no matter how this limit is taken, upon modular transformation to fundamental domain of $\Omega$, are known to densely populate it with the canonical $\Sp(2g,\R)$-invariant measure \cite{clozel2001hecke}.  
The average over the unitary group $\U (g)$ emerges because of the pseudo-random unitaries $U(\gamma,\Omega)$ generated by these modular transformations. As a result the average is over $\Sp(2g,\Z)\backslash \Sp(2g,\R)$, which is the full moduli space in 4d.  

The average is finite only for $n\leq g$. In particular for $g=n=1$ it is given by 
\bea
\label{ng=1}
\langle \tilde\theta(\tau,r,z e^{i\varphi}, \bar z e^{-i\varphi})\rangle_{\tau,\varphi}=r\, e^{\pi \bar z^T z}+r^{-1}e^{-\pi \bar z^T z},
\eea
where $G=r^2$. 

Similar considerations for orthogonal group and prime $N$ yields 
\bea
\label{LHSC}
{\Phi\over \NC} \sum_\CC\langle {\cal B}|\CC\rangle={N^{ng/2}\over \NC}\sum_{h\in G_O}{\sf \Psi}_{0\dots 0}(\Omega,E_h,z_h)={1\over \NC}\sum_\CC \tilde\theta(\Omega,E_\CC,z), \\
G_O=\Gamma_0\backslash O(n,n,\Z_N),\quad \NC=|G_O|.
\eea
Again, the sum over $E_\CC$ can be represented via Hecke operator for the orthogonal group. 
When $n>g+1$, in the $N\rightarrow \infty$ limit the sum converges, conjecturally leading to an average over the full Narain moduli space $\Or (n,n,\R)/O(n,n,\Z)$,
\bea
{1\over \NC} \sum_{\CC} \tilde\theta(\Omega, E_\CC, z,\bar z)=\langle \tilde\theta(\Omega,E,O_L z, O_R \bar z )\rangle_{E,O_L,O_R}.
\eea
We expect this conclusion to be true when $N\rightarrow \infty$ independently of whether $N$ is prime. 
In the opposite case of $n<g+1$ when $N$ is prime  we find 
\bea
\label{EisO}
\sum_{\gamma \in G_S} {\sf \Psi}_{0\dots 0}(\Omega, E_h,\xi_h)={\rm det}(G)^{g/2} E^{O}_g(E,z),\qquad \xi=z\,\Omega_2^{1/2},\\
E^{O}_g(E,z,\bar z)=\sum_{h \in \Gamma_0\backslash O(n,n,\Z)} { e^{\pi \bar z^T O_R^T O_L z}\over |{\rm det}(C E^T+D)|^g}.
\eea
Matrices $O_L,O_R$ are defied in terms of $h,E$ as discussed below \eqref{orthogonaltransformation}.
In particular for $n=1$, c.f.~\eqref{ng=1}, $E^{O}_g(E,z,\bar z)=r\, e^{\pi \bar z^T z}+r^{-1}e^{-\pi \bar z^T z}$, where $E=G=r^2$.

Combining all together,  in the $N\rightarrow \infty$ limit we find for $g>n-1$, i.e.~when the 4d manifold is fixed and the central charge of Maxwell theory is sufficiently large, 
\bea
\label{result}
\Phi_{5d} \langle Z_{\Omega}[\M_4]\rangle=\langle \tilde{\theta}(\Omega,E,z\, U)\rangle_{\Omega,U}=
{{\rm det}(G)^{g/2}} E^{O}_g(E,z,\bar z),\quad g>n-1.
\eea
In the same limit and when $n>g+1$ we recover the 3d result \cite{Afkhami-Jeddi:2020ezh,Maloney:2020nni,Datta:2021ftn}, where on the left is the average over 2d Narain theories,
\bea
\label{result3d}
\Phi_{3d}\langle Z_{E}[\Sigma]\rangle=\langle \tilde{\theta}(\Omega,E,O\,z)\rangle_{E,O}=
{{\rm det}(\Omega)^{n/2}} E_n(\Omega,z,\bar z),\quad n>g+1.
\eea
The second equality in \eqref{result} and \eqref{result3d} are the versions of the Siegel-Weil formula.  
In the derivation of \eqref{result} and \eqref{result3d} we assumed that $N$ is prime, but in both cases for any $N$ the LHS is the average over Hecke points leading to averages over corresponding moduli spaces, and hence we expect both of these relations to hold for $N\rightarrow \infty$ independently of how this limit is taken. 

The same expressions above  evaluate the divergent part of the 4d and 2d ensemble average in the  limit of large sqaure-free $N$ when the genus is larger than the central charge: 
\bea
\Phi_{5d}\langle Z_{\Omega}[\M_4]\rangle_{\Omega}=N^{-gn/2}\prod_k \prod_{i=1}^{g}{(p_k^i+p_k^n)\over (p_k^i+1)}  {{\rm det}(\Omega)^{n/2}} E_n(\Omega,z,\bar z),\quad n>g+1,
\eea
and 
\bea
\label{singular2}
\Phi_{3d}\langle Z_{E}[\Sigma]\rangle_{E}=N^{-gn/2}\prod_k \prod_{i=0}^{n-1}{(p_k^i+p_k^g)\over (p_k^i+1)} {{\rm det}(G)^{g/2}} E^{O}_g(E,z,\bar z),\quad g>n-1.
\eea
To derive these expressions we used \eqref{Nprime} and \eqref{5dA} which  assumme $N=\prod_k p_k$ is square-free. 
Coefficients in the RHS are singular when $N\rightarrow \infty$  scaling as $N^{g(n-1-g)/2}$ and $N^{n(g+1-n)/2}$ correspondingly, but the proportionality coefficient depends on the way $N$ is taken  to infinity. 

Finally we consider the case of non-square-free $N=p^2$ for a prime $p\rightarrow \infty$, which is discussed in detail for arbitrary $n$ and $g=1$ in Appendix \ref{matchingRHS}. In 5d, we find $\tau_\LL$ to be given by the $\NL=p^2+p+1$ Hecke points
\bea
\tau_{0}=p^2\tau,\qquad \tau_{{r+1}}={\tau+r\over p^2},\, r=0\dots p^2-1,\qquad \tau_{p^2+1+r}=\tau+{r\over p},\, r=0\dots p-1. \nonumber
\eea
Their average in the $p\rightarrow \infty$ limit is the average over the fundamental domain of $\tau$, which converges for $n=1<g+1$. In this case  we find \eqref{g=n=1}, which in the large $p$ limit reduces to \eqref{result}. Similarly in the 3d case, average over $\NC(N=p^2)$ Narain theories converges converges for $n>2$, in which case the holographic identity is given by \eqref{3dp2}. In the large $p$ limit the contribution of in the RHS $\LL'$ vanishes -- that is a particular Narain theory characterized by $E$ with a vanishing weight coefficient -- while the Poincare series of vacuum converges to the Eisenstein series. Hence, again, \eqref{result3d} emerges in the $N=p^2\rightarrow \infty$ limit. 
At the same time considering $g=n=1$ in 3d, we find that in the large $N=p^2\rightarrow \infty$ limit the average over $\NC=3$ Narain theories diverges, the singular term  is correctly given by $N^{1/2}{{\rm det}(G)^{g/2}} E^{O}_g(E,z,\bar z)$, but the proportionality coefficient is different from the one in \eqref{singular2}.

To summarize, in the large $N$ limit, in both 2d and 4d, whenever the ensemble over filed theories converges, it is given by the Eisenstein series, which holographically can be interpreted as the sum over 3d and 5d handlebody geometries. Thus $N\rightarrow\infty$ is a ``semiclassical'' limit when the singular geometries can be omitted from the bulk sum. 

When the field theory average diverge, the leading singularity is given by the appropriate Eisenstein series (of orthogonal group in 2d and sympathetic group in 4d), but the overall coefficient is not universal, e.g.
\bea
\langle Z_{\Omega}[\M_4]\rangle_{\Omega}\propto N^{n(g+1-n)/2}{{\rm det}(\Omega)^{n/2}} E_n(\Omega,z,\bar z),\quad n>g+1.
\eea

In the discussion above we assumed that $n>g+1$ or $g>n-1$. When $n=g+1$, in the large $N$ limit both sides of holographic identity diverges. The case of $n=g+1=2$ was  analyzed in detain in \cite{Aharony:2023zit}. It would be interesting to extend this analysis to arbitrary $n=g+1>2$.

\subsection{Correlators of local operators}
\label{sec:correlators}
\begin{figure}
    \centering
    \includegraphics[width=1.0\linewidth]{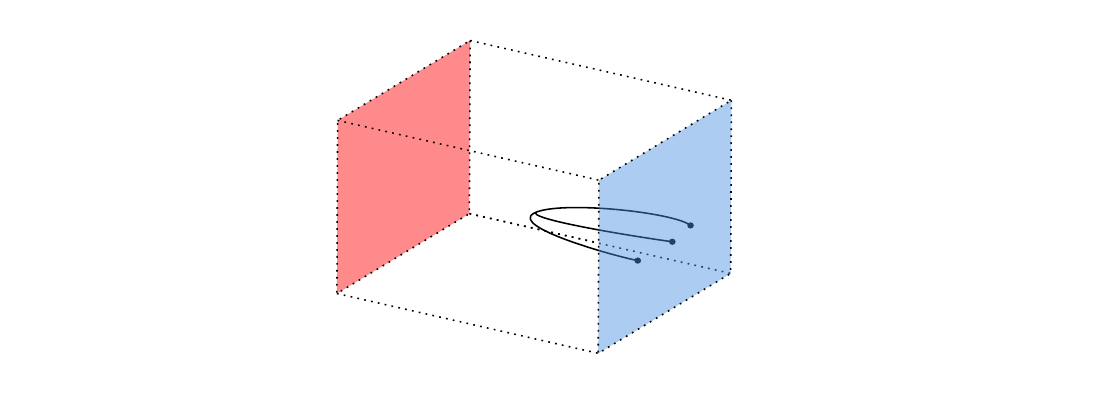}
    \caption{2d CFT conformal block with the operator insertions, given for Abelian theory by \eqref{conformalblockA}, is given by TQFT on $\Sigma\times [0,1]$ with the line operators (Wilson lines) ending at the conformal boundary (depicted in blue). }
    \label{fig:WL}
\end{figure}

We start with a general discussion of correlators in the context of 2d/3d (R)CFT/TQFT correspondence. In this case a conformal block involving primary operators $V_i(w_i)$ characterized by quantum numbers $h_i,\bar h_i$ is evaluated by the 3d path integral on $\Sigma \times [0,1]$ with the line operators ending at the points of operator insertions $w_i$. 
When boundary is a torus, this can be written explicitly as 
\bea
\label{CB}
\chi^{h_i}_c(\tau,w_i)={\rm Tr}_{{\cal H}^{2d}_{c}}\left(e^{-\beta H}\prod_i V_i(w_i)\right).
\eea
The bulk interpretation of this expression is depicted in Fig.~\ref{fig:WL} where the TQFT state at the ``inner'' (pink) boundary is chosen to be $|c\rangle \in \HH$, corresponding to the sector of the 2d CFT Hilbert space ${\cal H}_{2d}^{c}$ traced over in \eqref{CB}. Since the line operators do not end at the inner boundary, by keeping the boundary conditions there arbitrary, the bulk path integral can be understood as a ``boundary'' state $\langle {\cal B}(h_i,w_i)|$ in the dual to the Hilbert space  $\HH^g$ of the TQFT on $\Sigma \times \R$,  
\bea
\chi^{h_i}_c(\Omega,w_i)=\langle {\cal B}(h_i,w_i)|c\rangle.
\eea
Choosing instead the topological boundary conditions $\CC$ at the inner boundary would yield the correlator in the 2d CFT specified by $\CC$,
\bea
\langle \prod_i V_i(w_i)\rangle_\CC \, Z_\CC(\Omega)=\langle {\cal B}(h_i,w_i)|\CC\rangle.
\eea

We note that that  in general $\chi^{h_i}_c(\Omega,w_i)$ and even $\langle \prod_i V_i(w_i)\rangle_\CC$
are not necessarily  single-valued function of the insertion points $w_i$. The latter would require $V_i$ to be local with respect to $c$, or be present in the spectrum  of $Z_\CC$. In this case line operators can be easily pulled into, and back from, the inner boundary without any ambiguity. 

The statement of ensemble holography discussed in section \ref{sec:holography} ensures that the sum over topologies is equal to weighted average of topological boundary conditions 
\bea
\label{holcor}
\sum_\CC \alpha_\CC |\CC\rangle=|\Psi_{\rm TQFT\, gravity}\rangle. 
\eea
Weights $\alpha_\CC$ are positive and have probabilistic interpretation. 
Evaluating the scalar product of \eqref{holcor} with $\langle {\cal B}(h_i,w_i)|$ we immediately find that the statement of TQFT gravity being dual to the boundary ensemble extends to include correlators of primary operators, as was mentioned without explanation in 
\cite{Dymarsky:2024frx}.  

The fact that the equality between bulk and boundary partition functions on any $\Sigma$ \eqref{ensembleholography} extends to include correlators is not that surprising. A CFT partition function on a higher-genus Riemann surface ``knows'' about the correlation functions, which can be probed at the singular points of the moduli space ${\cal M}_g$ when $\Sigma$ factorizes. So far the holographic equality holds for any $\Omega$, it is natural to expect that it includes the correlators of primaries. 

Two comments are in order. First, the boundary ensemble includes exactly the same theories no matter we evaluate the partition function with or without the operator insertions. The operators $V_i$ may not be local for some of the theories in the ensemble, hence the resulting correlator will have brunch cuts. Second, the sum over topologies includes exactly the same terms independently of  $V_i$. Thus, if the sum includes only the handlebodies and hence can be written as an average over the mapping class group of $\Sigma$, which is e.g.~the case for the theory \eqref{AB} with a square-free $N$ \cite{Aharony:2023zit,Dymarsky:2025agh}, the same will be true after $V_i$ insertions. An example of such a calculation can be found in \cite{Benjamin:2021wzr}.
This is different from the proposal of \cite{Romaidis:2023zpx} which includes the average over the mapping class group of $\Sigma$ with punctured points $w_i$. We discuss this  difference in more detail below. 

We illustrate the discussion with an example of a general Abelian 3d CS theory  \eqref{3d} assuming for simplicity  $c_L=c_R$. (The result can be easily extended to any theory admitting topological boundary conditions.) The matrix $K_{ij}$ is the Gram matrix of some lattice $\Lambda$, while the line operators (Wilson lines) are labeled by the elements of the discriminant group $c\in \Lambda^*/\Lambda$. Topological boundary conditions (even self-dual codes) $\CC$ parameterize all possible Narain lattices $\Lambda_\CC$ satisfying $\Lambda \subset \Lambda_\CC \subset \Lambda^*$ \cite{Barbar:2023ncl}.  Wilson lines that end at the boundary are not subject to identifications modulo elements of $\Lambda$ and are parameterized by arbitrary vectors $\vec{k}=(\vec{k}_L,\vec{k}_R)\in \Lambda^*$ that play the role of quantum numbers $h,\bar h$. In fact corresponding vertex operators $V_i=:e^{i kX}:$ have conformal left and right conformal weights $h=k_L^2/2$, $\bar h=k_R^2/2$. 

Without any insertions the wavefunctions of the 3d theory quantized on $\Sigma\times R$ (non-analytic conformal blocks of 2d Narain CFT) are given by sums over shifted lattices $\Lambda_{c}$,  which includes all vectors of the form $\vec{v}+\vec{c}$, for all $\vec{v}\in \Lambda$ \cite{Aharony:2023zit,Barbar:2023ncl},
\bea
\chi_c(\Omega,\xi,\bar\xi)=\langle \Omega,\xi,\bar\xi|c\rangle.
\eea
For the AB theory \eqref{AB} these are the wavefunctions \eqref{3dblock}.
Focusing on the case of a torus, with the vertex operator insertions the block takes the form \cite{Dijkgraaf:1987vp,Mason:2002cg}
\bea
\label{conformalblockA}
\chi^{k_i}_c(\tau,z_i,\xi,\bar\xi)&=&\langle \tau,\xi,\bar\xi, V_i(w_i)|c\rangle\\ 
&=&\chi_c(\tau,\xi',\bar\xi')e^{-{\pi\over 2\tau_2} (w_L^2+w_R^2)}
\prod_{i<j} E(w_{ij}|\tau)^{k_L^i\cdot k_L^j} \overline{E(w_{ij}|\tau)}^{k_R^i\cdot k_R^j},
\nonumber \\ \nonumber
\xi'=\xi+w_L,\quad \bar\xi'&=&\bar \xi+w_R,\quad w_{L}\equiv\sum_{i}k_{L,i}\, w_{i},\quad  
w_{R}\equiv\sum_{i}k_{R,i}\, \bar{z}_{i},\quad w_{ij}\equiv w_i-w_j\\
\label{torusblock}
\chi_c(\tau,\xi,\bar\xi)&=&{\sum_{(p_L,p_R)\in \Lambda_c}e^{i\pi \tau p_L^2-i\pi \bar\tau p_R^2+2\pi i(p_L\xi-p_R\bar\xi)+\pi(\xi^2+\bar\xi^2)/\tau_2}\over |\eta(\tau)|^{2n}}.
\eea
Here $w_i$ parameterize the points on the torus, where the vertex operators are inserted and $\sum_i k_i=0$ lest the correlator vanishes. 
A simple dependence on $c$ in \eqref{conformalblockA} is a specific  feature of the Abelian theory.
The function \bea
E(w|\tau)={\theta_1(w|\tau)\over \theta_1'(0|\tau)}
\eea 
is called the prime form and its related to the Green's function  of the scalar Laplacian,  $G(w)=-\ln |E(w|\tau)| + \frac{2\pi\left({\rm Im}(w)\right)^{2}}{\tau_{2}}$ \cite{Polchinski:1998rq}.


Once  the ensemble-averaged correlator of $\prod_i V_i(w_i)$ is evaluated, 
\bea
\sum_\CC \langle \prod_i V_i(w_i)\rangle_\CC Z_\CC(\tau,\xi,\bar\xi)=\langle \tau,\xi,\bar\xi, V_i(w_i)|\Psi_{\rm TQFT\, gravity}\rangle,
\eea
the correlators of  currents, and hence $\U (1)$ descendants,  can be probed by considering a kinematic limit of $w_i$ merging pair-wise. 

We note that since   $V_i$ has positive conformal dimension, modular transformations will introduce a scalar prefactor, c.f.~\eqref{modtrbra},
\bea
&&\langle \tau,\xi,\bar\xi, V_i(w_i)|U^\dagger_\gamma=(c\tau+d)^{h}(c\bar\tau+d)^{\bar h}
\langle \tau_\gamma,\xi_\gamma,\bar\xi_\gamma, V_i(\gamma w_i)|,\\
&&\tau_\gamma={a\tau+b\over c\tau+d},\quad \xi_\gamma={\xi\over c\tau+d},\quad \bar\xi_\gamma={\bar\xi\over c\bar\tau+d},\quad \gamma w_i={w_i\over c\tau+d},\\
&& h={1\over 2}\sum_i (k^i_L)^2,\quad \bar h={1\over 2}\sum_i (k^i_R)^2.
\eea
Now, whenever the averaged 2d partition function is given just by the handlebody contributions, i.e.~by the vacuum Poincare series, 
\bea
\sum_\CC Z_\CC(\tau,\xi,\bar\xi)=\kappa \sum_{\gamma\in \Gamma_0\backslash SL(2,\Z)} \chi_0(\tau_\gamma,\xi_\gamma,\bar\xi_\gamma),
\eea
the same will apply to the correlators, with the Poincare series taking form 
\bea
\label{Poincare}
\sum_\CC \langle \prod_i V_i(w_i)\rangle_\CC Z_\CC(\tau,\xi,\bar\xi)=\kappa \sum_{\gamma\in \Gamma_0\backslash SL(2,\Z)} (c\tau+d)^{h}(c\bar\tau+d)^{\bar h}\chi^{k_i}_0(\tau_\gamma,\gamma w_i,
\xi_\gamma,\bar\xi_\gamma).
\eea
Note that the sum must explicitly include the pre-factor $(c\tau+d)^{h}(c\bar\tau+d)^{\bar h}$ because the character $\chi_0^{k_i}$ should be understood as a section of a line bundle. 

If instead of the mapping class group of the torus $SL(2,\Z)$ the vacuum character $\chi^{k_i}_0(\tau_\gamma,\gamma w_i,\xi_\gamma,\bar\xi_\gamma)$ were to be averaged over the mapping class group of the torus with punctured points, the result would be different. Namely, the boundary ensemble will only include theories for which all operators $V_i$ are local. 
More generally, as in the case without operator insertions, the consistent procedure would be to 
average over the mapping class group of the punctured Riemann surface $\Sigma$, starting from an infinitely large genus and then obtaining the result for finite $g$ via genus reduction. This  would lead to an average over the boundary ensemble 
that includes only theories for which all $V_i$ are local. The relative weights $\alpha_\CC$, which are non-trivial in the non-Abelian case,  would remain the same as in the case without insertions \cite{Barbar}.


Extending this picture to 4d/5d is straightforward. Surface operators of the $B_2$ and $C_2$ fields in the bulk with charges ${\rm k}$ -- vectors in the symplectic lattice \eqref{symplecticlattice} --   ending
on a closed but not necessarily connected  contour $\Gamma=\cup_i \Gamma_i$ at the boundary will be calculating correlators of the Wilson lines $W_i^{{\rm k}_i}$ of the gauge field $A_\mu$ and its dual $A_\mu^D$  over $\Gamma_i$. The conformal blocks with Wilson operator insertions 
\bea
\langle E,W_i^{{\rm k}_i}(\Gamma_i),\xi,\bar\xi|{\rm c}\rangle
\eea
will have a form similar to \eqref{conformalblockA}, namely, a conformal block without insertions multiplied by a ${\rm c}$-independent factor that depends on the contour $\Gamma$ and the Wilson lines charges.  It can be written explicitly in terms of the Green's function for a vector Laplacian on $\M_4$.
Similarly to the kinematic limit of $w_i$ merging pairwise, taking the boundary contour $\Gamma_i$ to encircle a small region will lead to a correlation function with   $F_{\mu\nu}$ or $F^D_{\mu\nu}$ in the pre-exponent.

\section{4d ${\cal N}=4$ SYM}
\label{sec:N=4}
In view of the ensemble holography picture developed above, an obvious question is whether similar results might apply to the case of 4d ${\cal N}=4$ SYM. We postpone this discussion until section \ref{sec:discussion}, while here we focus on a different question. Namely, which ${\cal N}=4$ SYM theory --- $\U (N)$  or ${\rm SU} (N)$ --- is IIB String Theory on $AdS_5 \times \SSS^5$ dual to? 
This question has been extensively discussed in the literature,
with strong arguments in favor of both the $\SU (N)$ \cite{Aharony:1998qu,Witten:1998wy,aharony2013reading,Kapustin:2014gua} and $\U (N)$ \cite{Maldacena:2001ss,belov2004conformal} scenarios. More recent works mention that both scenarios are possible  without providing  details \cite{Aharony:2016kai}. As we explain below,  the choice of gauge group is specified by the boundary conditions for the bulk fields $B_2, C_2$. 

We begin our analysis with the SymTFT construction for the family of ${\cal N}=4$  field theories with gauge algebra $\SU (N)$ on a simply connected spin four-manifold $\M_4$,  coupled to the 5d theory \eqref{5d} on $\M_4\times \R$ \cite{Witten:1998wy, gaiotto2015generalized, Apruzzi:2021nmk,Bergman:2022otk}. This coupling defines a particular boundary state $\langle {\cal B}_{\SU (N)}|$  such that 
\bea
\label{suNCB}
Z^{\SU (N)}_{ab}=\langle {\cal B}_{\SU (N)}|ab\rangle_{5d} 
\eea
are the conformal blocks of the 4d theory. These are partial sums over the 4d QFT Hilbert space, 
that includes only states with  particular values of electric and magnetic charges. With a proper $N$-dependent normalization the latter are vectors in the weight lattice of $su(N)$. Values of $a,b\in \Z_N$ specify the charges modulo vectors of the root lattice \cite{Kapustin:2005py,aharony2013reading}. As was discussed in section \ref{sec:SymTFT},  partition functions of all possible  theories with the $su(N)$ gauge algebra and different gauge groups are given by the topological boundary conditions of the SymTFT, 
\bea
\label{suNZ}
Z^{\SU (N)}_\LL=\langle {\cal B}_{\SU (N)}|\LL\rangle=\sum_{{\rm c}\in \LL} Z^{\SU (N)}_{\rm c},\qquad {\rm c}\in (\Z_N\times \Z_N)^n.
\eea 

Similarly, the SymTFT construction for ${\cal N}=4$ $u(1)$ theory is a slight generalization of the Maxwell theory case considered above. The corresponding conformal blocks
\bea
\label{susyu1}
Z^{u (1)}_{ab}=\langle {\cal B}_{u (1)}|ab\rangle_{5d} 
\eea
are essentially the same as in Maxwell theory (\ref{confblocks},\ref{5dtheta}), modulo the 
contributions of the superpartners. These are sums over states with  electric and magnetic charges, which after proper normalization are equal to $a,b$ mod $N$. The analog of \eqref{suNZ} are the partition functions of ${\cal N}=4$ $U(1)$ theories with different values of coupling as given by \eqref{tauL} and its generalizations. 

In what follows we will specify the gauge algebra when we want to emphasize that we consider all  conformal blocks of the form (\ref{suNCB},\ref{susyu1}). The gauge group $\U(1)$ or ${\rm SU}(N)$ will refer to a particular absolute theory, i.e.~a particular combination of these conformal blocks. 

\begin{figure}
    \centering
    \includegraphics[width=1.0\linewidth]{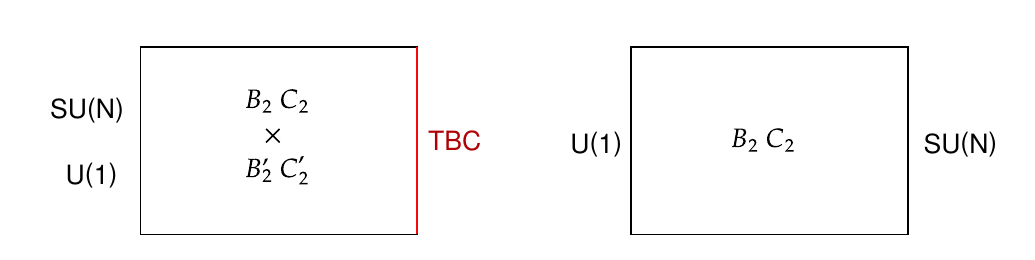}
    \caption{Gauging the center $\Z_N$ of $\U (1)\times {\rm SU} (N)$ theory in terms of SymTFT construction.}
    \label{fig:gauging}
\end{figure}

The field theory with gauge group $\U(N)$ can be understood in terms of $u(1)$ and $\SU (N)$ theories as follows. 
At the group level $\U (N)=({\rm SU} (N)\times \U (1))/\Z_N$, hence the $\U(N)$ theory is a basically a product $\U(1)$ and ${\rm SU} (N)$ theories with the diagonal $\Z_N$ gauged. The resulting theory can described in terms of a SymTFT construction as follows. We consider the $\U(1)$ theory with coupling $\tau_0=N\tau$ on $\M_4$ and couple it to the SymTFT \eqref{5d} with $g=1$ and the fields $B_2,C_2$ living in $\M_4\times [0,1]$. The resulting wavefunctions in the bulk are the conformal blocks of \eqref{susyu1}.
We then consider the ${\rm SU} (N)$  theory with coupling $\tau'$  on $\M_4$ and couple it to another SymTFT  on $\M_4\times [0,1]$ with fields $B_2',C_2'$. The corresponding wavefunctions are as in \eqref{suNCB}.
So now the bulk includes two copies of $\Z_N$ gauge theory, with $B_2,C_2$ in the fundamental and $B_2',C_2'$ in the anti-fundamental representation of $SL(2,\Z)$. To gauge the center $\Z_N$ we cap the cylinder $\M_4\times [0,1]$ with the topological boundary condition corresponding to the diagonal invariant. This is depicted in the left panel of Fig.~\ref{fig:gauging}. 
Using the (un)folding trick, an equivalent construction can be obtained by placing the $U(1)$ and ${\rm SU} (N)$ field theories on opposite ends of the cylinder, with just a single $B_2,C_2$ theory living inside. This arrangement is depicted in the right panel of Fig.~\ref{fig:gauging}, and it
leads to the following expression for the partition function after gauging 
\bea
\label{UN}
Z_{\U (1)\times {\rm SU} (N)\over \Z_N}=
\sum_{a,b\in \Z_N^n} \langle {\cal B}^\tau_{u (1)}|a,b\rangle \langle {\cal B}^{\tau'}_{\SU (N)}|a,b\rangle =\sum_{a,b\in \Z_N^n} \langle {\cal B}^\tau_{u (1)}|a,b\rangle \langle a,b|{\cal B}^{-\bar \tau'}_{\SU (N)}\rangle. 
\eea
In the expression above,  instead of the $\SU (N)$ conformal blocks with coupling $\tau'$ we used the conjugate conformal blocks with coupling $-\bar \tau'$. Taking $\tau=-\bar \tau'$ will yield the  conventional ${\cal N}=4$  $\U (N)$ gauge theory
\bea
Z_{\U (N)}=\sum_{a,b\in \Z_N^n} {Z}^{u(1)}_{ab}\,\bar{Z}^{\SU(N)}_{ab}.
\eea
Distinct $\tau\neq -\bar\tau'$ corresponds to an exactly marginal deformation.

We now turn to the bulk description, and consider the low energy limit of IIB supergravity on $X_{\rm bulk}\times \SSS^5$, where $X_{\rm bulk}$ behaves as $AdS_5$ near the boundary $\M_4=\partial X_{\rm bulk}$. In this limit, at the level of the bulk action the topological sector described by \eqref{5d} decouples from the other fields \cite{Witten:1998wy}. At the same time, the extended objects -- the fundamental string F1 and the D1-brane that are holographically dual to Wilson line and 't Hooft line operators respectively \cite{Maldacena:1998im,Rey:1998ik,Kapustin:2005py} -- are still charged under $B_2$ and $C_2$ correspondingly. Assuming that $X_{\rm bulk}$ is topologically the same as the
handlebody $X_5$ used to define the basis $|a,b\rangle_{5d}$, the semiclassical gravity configuration dual to the conformal block $ \bar Z_{ab}^{\SU (N)}$ will include F1 and D1 branes wrapping $n$ two-cycles of $\M_4$ that are not shrinkable inside $X_{\rm bulk}$,  such that the numbers of F1s and D1s are  equal to $a$ and $b\,\,{\rm mod}\,\, N$. If  the two-cycles of $\M_4$ shrinkable inside $X_{\rm bulk}$ and $X_5$ are not the same, these configurations will contribute to a conformal block $\langle a,b|{\cal B}_{\SU (N)}\rangle$ evaluated in a different basis. The boundary partition function will include a sum over all $a,b$ and the result will be basis-independent.
If there are several semiclassical bulk geometries $X_{\rm bulk}$ satisfying the supergravity equations of motion that can end on $\M_4$, which is often the case \cite{Aharony:2024fid}, the contributions of all such $X_{\rm bulk}$ should be summed over.  

The important point, since the F1- and D1-branes are charged under $B_2$ and $C_2$, is that the  semiclassical bulk calculation will automatically evaluate 
\bea
\sum_{a,b\in \Z_N}\langle {\cal B}|ab\rangle_{5d}\, \bar Z_{ab}^{\SU (N)},
\eea
where the state $\langle {\cal B}|$ depends on the boundary conditions (the quantization scheme) for the bulk fields $B_2,C_2$. 
This is in fact completely in parallel with the SymTFT construction discussed above. The boundary conditions for the  topological sector define the boundary state ${\cal B}$ and the topological fields $B_2,C_2$ ``live'' in the near-boundary region, while the rest of the fields and branes inside the bulk are combined into ${\cal B}_{\SU (N)}$. The resulting picture is the same as in the right panel of Fig.~\ref{fig:gauging}, as  discussed in detail in \cite{Bergman:2025isp}.

Assuming we quantize the  topological sector via the holomorphic quantization prescription of section \ref{sec:holquantization}, e.g.~by adding corresponding boundary terms  as was suggested in \cite{Maldacena:2001ss}, complemented by ${\cal N}=4$ SUSY-preserving boundary conditions for the superpartners,\footnote{
The bosonic superpartners of the $\U (1)$ gauge field are 6 scalars $\phi_I$. In $AdS_5\times \SSS^5$ they are dual  to non-normalizable modes  of the warp factor $h$, $ds_{10d}^2=h^{-1/2}dx_\mu^2+h^{1/2} dX_I^2$. The latter satisfies $\nabla_X^2 h=-(2\pi)^4 g_2 \alpha'^2 N \delta^6 (X)$. The corresponding modes are due to the collective movement of the D3 branes, that shift the location of the delta-function singularity. } the boundary state will be ${\cal B}_{u(1)}$ from \eqref{susyu1}. In this case, the semiclassical bulk partition function, which includes a sum over all $a,b$ sectors, 
\bea
\label{Zbulk}
Z_{\rm bulk}=\sum_{a,b\in \Z_N^n} \langle {\cal B}_{u (1)}|ab\rangle_{5d}\, \bar Z_{ab}^{\SU (N)}
\eea
will evaluate the partition function of the $\U (N)$ theory \eqref{UN}, as was pointed out in 
\cite{belov2004conformal}.

We note that the value of the $u(1)$ gauge field coupling constant $\tau$ will be fixed by the value of the boundary term \eqref{5dboundaryterm}, and should be tuned to agree with the coupling constant of the $\SU (N)$ gauge theory, which is determined by the value of the axion-dilaton bulk fields near the boundary \cite{Aharony:1999ti}.  Another way to look at the resulting construction is to say that the boundary terms describes a dynamical 4d $\U (1)$ theory, which is coupled to IIB supergravity via the topological $B_2,C_2$ fields \cite{Aharony:2016kai}.\footnote{We thank O.~Aharony for discussions on this point. }  

Alternatively one can impose topological boundary conditions on the bulk fields at $\M_4=\partial X_{\rm bulk}$, as discussed in section \ref{sec:SymTFTpartI}.
In such  a case $\langle {\cal B}|=\langle \LL|$ and \eqref{Zbulk} will reduce to $Z_{\rm bulk}=Z^{su(N)}_\LL$. In other words, imposing a particular topological boundary condition will result in the bulk being dual to a particular theory with  $su(N)$ gauge algebra. 

Thus, depending on the boundary conditions for the  fields $B_2,C_2$ in the bulk, e.g.~\eqref{selfdual} or \eqref{topbc}, or equivalently on the quantization scheme (i.e.~the choice of polarization -- holomorphic or real), IIB supergravity in the bulk will be dual to the $\U (N)$ theory or to a particular $su (N)$ theory at the boundary.

The above discussion was in the low-energy limit, in which the topological sector decouples at the level of the bulk action. 
If we now include corrections of higher order in $\alpha'$, the two-form fields $B_2,C_2$ will acquire kinetic terms and the corresponding sector will no longer be strictly topological. As was shown in \cite{belov2004conformal}, the 
low-energy limit of this theory (without any additional boundary terms) is equivalent to the topological theory \eqref{5d} in {\it holomorphic} quantization. In this case the value of $\tau$ entering the explicit form of the bulk wavefunctions (\ref{3dtheta}, \ref{5dtheta}) is specified by the kinetic terms, and will automatically be the same as the value of the axion-dilaton at the boundary. Hence the most straightforward quantization of IIB supergravity yields the gravity dual of the $\U (N)$ theory \cite{Maldacena:2001ss,belov2004conformal}.  Yet even if higher derivative corrections are included, or even in the full IIB String Theory, all the other scenarios we have described are still possible. We note that imposing e.g.~$B_2=0$ at the boundary when kinetic terms are present will only enforce ${\rm Im\,}\xi={\rm Im\,} \bar \xi=0$ within holomorphic quantization. Instead, to impose topological boundary conditions, one should take into account that the basis of states $\langle \tau,E,\xi,\bar \xi|$ within $\HH^*$ is over-complete; hence any state $\langle {\cal B}|\in \HH^*$ imposing appropriate boundary conditions is possible. In practice this means that to impose a topological boundary condition $\LL$, one would need to integrate the boundary values of the bulk field over $d\xi\, d\bar \xi$ against the kernel $\langle \LL|\tau,E,\xi,\bar \xi\rangle$. 

We end by noting that more exotic boundary conditions are possible. Consider for example the ${\cal N}=4$ SYM theory on $\M_4$, or equivalently IIB String Theory in the bulk ending on $\M_4$, coupled to an {\it auxiliary} 5d topological theory \eqref{5d} with fields $\tilde{B}_2,\tilde{C}_2$. This auxiliary theory will be living in an ``auxiliary'' 5d bulk, so the resulting construction has two 5d bulks -- the conventional one and the auxilary one 
ending on the same 4d manifold $\M_4$, where we require $\tilde{B}_2=B_2, \tilde{C}_2=C_2$.\footnote{A similar construction with two bulks was recently discussed in \cite{Heckman:2025lmw}. There, in contrast with our setup, the topology of the auxiliary bulk is fixed.} Now if we sum over all possible topologies of the auxiliary bulk, thus promoting the auxiliary 5d theory to a gravitational TQFT, then as explained in section \ref{sec:holography} the resulting state of the $\tilde{B}_2,\tilde{C}_2$-theory will be a sum $\sum_{\LL}|\LL\rangle$ over all topological boundary conditions. This means that our construction with two bulks is
holographically  dual to an ensemble of all ${\cal N}=4$ $su(N)$ theories

\section{Conclusions}
\label{sec:discussion}
In this paper we have formulated a holographic duality between a model of 5d topological gravity --- 5d Abelian TQFT \eqref{5d} summed over all 5d topologies sharing the same boundary -- and an ensemble of 4d Maxwell theories living on the boundary manifold. 
The duality implies the equivalence of partition functions, as well as of correlators of local  operators, primary with respect to the symmetry algebra defined by the bulk TQFT,
as discussed in section \ref{sec:correlators}.   
This duality is a direct extension of the duality between the 3d Abelian Chern-Simons theory with compact gauge group summed over topologies and an ensemble of 2d Narain CFTs at the boundary \cite{Aharony:2023zit,Dymarsky:2025agh}. 
In the large $N$ limit we find that the boundary theories densely cover the space of gauge couplings with the canonical measure. The average partition function is well-defined when the 4d central charge is sufficiently large and is given by the Eisenstein series of the orthogonal group $\Or (n,n,\Z)$. This is a version of the Siegel-Weil formula which made an appearance previously in the  2d/3d context  \cite{Afkhami-Jeddi:2020ezh,Maloney:2020nni,Datta:2021ftn} and many years earlier in the context of multiloop string amplitudes \cite{Obers:1999es,Obers:1999um}. The bulk sum over topologies in 
the $N\rightarrow \infty$ limit 
 includes only handlebody geometries, suggesting the bulk theory becomes semiclassical. 
We have also shown that the holographic duality of both the 4d/5d and 2d/3d cases extends to correlators of $\U (1)$-primaries and their descendants, as discussed in section \ref{sec:correlators}.


The established lore  suggests that  ensemble holography is a feature of lower dimensions when there is no dynamical graviton.
Our bulk theory is a topological theory of gravity and we see  no qualitative difference between  lower and higher dimensions. In our setting both scenarios, with a boundary ensemble and with a unique boundary theory, are equally valid. Which of these two scenarios is realized is determined by the properties of the bulk TQFT.  If the bulk theory is topologically trivial, corresponding to level $N=1$, the boundary ensemble includes only one theory. 

To bridge the gap between topological and conventional gravity in the bulk, 
we need to address two different but related questions. The first question is how to extend our setup to include a dynamical graviton in the bulk. The second is to understand how our setup connects to the semiclassical gravity regime.  In what follows we focus on the 2d/3d case, which is simpler. 

With regard to the first question, an obvious limitation of our setup is that the classical sources $J$ in \eqref{holography} couple only to boundary currents of the $\U (1)^n\times \U (1)^n$ symmetry, while the stress-energy tensor is built out of these currents by the Sugawara construction. To introduce classical sources for $T_{\mu\nu}$ directly, the first step would be to extend the bulk TQFT to include a line operator for each conformal primary. The resulting Virasoro TQFT (VTQFT) could potentially describe, in the sense of the sandwich construction, any 2d conformal theory. Accordingly, a gravitational theory that sums the VTQFT over all 3d topologies will be dual to a weighted unitary ensemble of all 2d CFTs. That is schematically the same as what happens in pure 3d quantum gravity, see \cite{Collier:2023fwi,Collier:2024mgv,Post:2024itb,Hartman:2025cyj,Hartman:2025ula,Chandra:2025fef} for related developments, although in that case the result is divergent and requires regularization. This divergence could be more than a technical problem --- different regularization schemes might lead to either scenario, with or without an ensemble, underlying the notion that both are equally valid. In order to break conformal symmetry and couple external sources to the stress tensor directly, one should go beyond VTQFT and introduce a line operator for each state in the boundary QFT Hilbert space.  

Another important question to address is the semiclassical gravity limit. In 3d we expect it to emerge ``automatically'' from the VTQFT summed over 3d topologies, in the limit of large central charge. In fact we have already seen an avatar of this behavior in the Abelian case, where for $N\rightarrow \infty$ the bulk sum includes only handlebody geometries, as in the semiclassical gravity case \cite{Maloney:2007ud}. This is in contrast to the trivial $N=1$ theory that does not differentiate between topologies. The sum over handlebodies as $N\rightarrow \infty$  does not imply, however, that the gravitational bulk theory admitting a semiclassical regime is necessarily dual to an ensemble. As we discussed in \cite{Dymarsky:2020pzc} and section 5 of \cite{Aharony:2023zit} the topological bulk theory dual to a single, {\it typical} Narain CFT can be recast in a form that includes a ``semiclassical'' sum over handlebodies, although this representation is 
far from 
unique. 

To summarize the discussion above, at this point we do not see any qualitative difference between lower and higher dimensional models of holography.  Both types of holographic duality, involving either a single boundary theory or an ensemble, can exist in higher dimensions. The ensemble interpretation seems more general; it reduces to the single-theory scenario for special choices of the bulk theory.

In addition to establishing the 4d/5d holographic duality between Maxwell theories and 5d Abelian TQFTs, in section \ref{sec:SymTFT}  this paper develops a connection between 5d Abelian TQFTs and codes.  We briefly summarize this connection here.
\begin{itemize}
\item Non-anomalous subgroups of the 2-form symmetry group in 5d are parameterized by classical symplectic codes. Maximal non-anomalous subgroups are in one-to-one correspondence with symplectic self-dual codes $\LL$. The TQFT states defined by topological boundary conditions are quantum stabilizer states $|\LL\rangle$ defined in terms of classical codes $\LL$ via the CSS construction.
This is an extension of the 3d story, where the underlying codes are even \cite{Barbar:2023ncl}.

\item Translating this result into the language of anyons (topological defects), and working in any number of dimensions, the Hilbert space of a theory obtained via (partial) anyon condensation in an Abelian TQFT is a quantum stabilizer code of CSS type, parameterized by a self-orthogonal classical code. The meaning of self-orthogonality, i.e.~the choice of the inner product, depends on the dimension. For 3d theories the corresponding codes are even, for 5d theories they are symplectic. 

\item Up to an overall normalization, the path integral of the 3d Abelian TQFT  \eqref{AB} on any 3d manifold with  boundary 
is a stabilizer state $|\LL\rangle$ specified by a classical symplectic self-dual code $\LL$, i.e.~a state of the 5d theory \eqref{5d} defined by a topological boundary condition. (Here we invoke the isomorphisms between the Hilbert spaces $\HH_\Sigma$ and $\HH_{\M_4}$ of the 3d and 5d theories.) Similarly, up to an overall normalization, the path integral of the 5d theory on any topology with boundary is a stabilizer state $|\CC\rangle$ defined by a classical even self-dual code $\CC$, which is also a state defined by a topological boundary condition in the 3d theory.  The intriguing relation between topologies in 3d/5d and topological boundary conditions in 5d/3d begs for a geometric interpretation in terms of the 7d 3-form theory. 
\end{itemize}

As a spin-off development we clarified the holographic dictionary between the gauge  group of 4d ${\cal N}=4$ SYM theory and the boundary conditions of the IIB String Theory fields $B_2,C_2$  in the bulk. As discussed in section \ref{sec:N=4},
holomorphic quantization of these fields, or equivalently the self-dual boundary condition \eqref{selfdual} and its generalization to non-zero $\xi,\bar\xi$, will yield the bulk dual of the $\U (N)$ theory. Imposing a topological boundary condition for $B_2,C_2$ instead will result in a particular $\SU (N)$ theory.

\appendix
\section{Modular and orthogonal transformations}
\label{Modular}
Modular transformations of $|(\alpha,\beta)\rangle_{3d}$ are  specified for the generators of the symplectic group $\Sp(2g,\Z)$ (mapping class group of $\Sigma$), mapped to $\Sp(2g,\Z_N)$,  
\bea
\gamma=\left(\begin{array}{cc}
A & B\\
C & D\end{array}\right)\in \Sp(2g,\Z_N)
\eea
that preserves the intersection matrix of one-cohomologies on $\Sigma$
\bea
J=\left(\begin{array}{cc}
0 & \mathbb{1}_g\\
-\mathbb{1}_g & 0\end{array}\right).
\eea
For invertible $A\in GL(g,\Z)$ and $D=(A^{-1})^T$, $B=C=0$ the transformation is simple 
\bea
U_\gamma |(\alpha,\beta)\rangle_{3d}=|A^{-1}(\alpha,\beta)\rangle_{3d},
\eea
where $A$ is acting on $\alpha_I$ and $\beta_I$ as on fundamental vectors. 

For $A=D=\mathbb{1}_g$, $C=0$ and integer symmetric $B$ -- a generalization of $T$-generator of $SL(2,\Z)$ -- action on basis elements is a pure phase
\bea
U_\gamma |(\alpha,\beta)\rangle_{3d}=e^{-{2\pi i \over N}\alpha^T B \beta}|(\alpha,\beta)\rangle_{3d}.
\eea

Finally, $\gamma=-J\in \Sp(2g,\Z)$ acts by the Fourier transform 
\bea
U_\gamma |(\alpha,\beta)\rangle_{3d}={1\over N^{gn/2}}\sum_{\tilde\alpha,\tilde\beta\in \Z_N^{gn}} 
e^{{2\pi i \over N}(\alpha \tilde\beta+\tilde\alpha \beta)}|(\tilde\alpha,\tilde\beta)\rangle_{3d}.
\eea

Similarly generators of the orthogonal group $\Or (n,n,\Z)$ 
that preserves the intersection form $\eta$  \eqref{eta} of two-cohomologies on $\M_4$ is mapped to 
\bea
\label{etaapp}
h=\left(\begin{array}{cc}
A & B\\
C & D\end{array}\right)\in \Or(n,n,\Z_N)  
\eea
which acts on $|(a,b)\rangle_{5d}$ as follows.
For invertible $A\in GL(g,\Z)$ and $D=(A^{-1})^T$, $B=C=0$ the transformation is simple 
\bea
U_{h} |(a,b)\rangle_{5d}=|(a,b)A^{-1}\rangle_{5d},
\eea
where $A$ is acting on $a_i$ and $b_i$ as on fundamental co-vectors. 

The generator $h=\eta\in \Or(n,n,\Z)$ acts by the Fourier transform 
\bea
U_h |(a,b)\rangle_{5d}={1\over N^{gn/2}}\sum_{a', b' \in \Z_N^{gn}} 
e^{{2\pi i \over N}{\rm Tr}(a^T b'-b^T a')}  |( a',b')\rangle_{5d}.
\eea

For $A=D=\mathbb{1}_n$, $C=0$ and integer antisymmetric $B$ action on the basis elements is a pure phase
\bea
U_h |(a,b)\rangle_{5d}=e^{-{2\pi i \over N}{\rm Tr}(a^T B b)}|(a,b)\rangle_{5d}.
\eea

\section{Quantization and the dimensional reduction of 7d theory}
\label{app:dimreduction}
\subsection{2d geometry and 3d Chern-Simons theory}
We start with geometric preliminaries. A Riemann surface $\Sigma$ of genus $g$ admits a basis of real-valued one-forms $\omega_I^{(1)}$, $I=1\dots 2g$ with the canonical intersection form 
\bea
\int_\Sigma \omega^{(1)}_I \wedge  \omega^{(1)}_J=J_{IJ},
\qquad 
J=\left(\begin{array}{cc}
0 & \mathbb{1}_g\\
-\mathbb{1}_g & 0
\end{array}\right),
\eea
such that first $g$ forms $\omega^{(1)}_I$ are dual to ``a''-cycles, and $\omega^{(1)}_{I+g}$ are dual to ``b''-cycles.
Next we introduce the metric 
\bea
\label{2dmetric}
\int_\Sigma \omega^{(1)}_I \wedge  \star\, \omega^{(1)}_J&=&{\mathbb G}_{IJ},\qquad \mathbb{G}={\mathbb\Lambda}^T {\mathbb\Lambda},\qquad
{\mathbb \Lambda}=\Omega_2^{-1/2}
\left(\begin{array}{cc}
-\Omega_1 & 1\\
\Omega_2 & 0
\end{array}\right),\\
\mathbb{G}&=&\left(\begin{array}{c|c}
\Omega_1 \Omega_2^{-1}\Omega_1+\Omega_2 & -\Omega_1 \Omega_2^{-1}\\ \hline
- \Omega_2^{-1} \Omega_1 & \Omega_2^{-1}\end{array}\right),
\eea
defined in terms of the modular  parameter $\Omega=\Omega_1+i\,\Omega_2$ of $\Sigma$. 
Here the Hodge star $\star$ is defined such that $\int A\wedge \star A$ is positive-definite for any one-form $A$, i.e.~the holomorphic differentials on $\Sigma$ will be the $-i$ eigenvectors of $\star$.
The metric \eqref{2dmetric} is compatible with the intersection form, i.e.~$J^{-1}\mathbb{G}=-\mathbb{G}^{-1}J$.

The holomorphic differentials can be written explicitly as follows, 
\bea
\label{real-hol}
\omega^{}_I=\omega^{(1)}_I+\Omega_{IJ}\, \omega^{(1)}_{g+J}, \quad I,J=1\dots g.
\eea
A straightforward calculation gives  
\bea
\label{2dint}
\int_\Sigma \omega_I^{} \wedge \omega_J^*=-2i\, (\Omega_2)_{IJ}.
\eea
The modular group 
\bea
\gamma=\left(\begin{array}{cc}
A & B\\
C & D\end{array}\right)\in \Sp(2g,\Z)
\eea
that acts fundamentally on the vector of cohomologies $(\omega^{(1)}_I,-\omega^{(1)}_{g+I})$ acts  
on $\Omega$ in the standard way 
\bea
\Omega\rightarrow (A\Omega+B)(C\Omega+D)^{-1}.
\eea

Now we consider chiral level-$N$ Chern-Simons theory 
\bea
\label{3dcs}
{N\over 4\pi}\int {\cal A}\wedge d{\cal A}
\eea
on $\Sigma\times \R$. This is an auxiliary theory not related to \eqref{3d}. To implement the 
holomorphic quantization we add the boundary term \cite{Aharony:2023zit}
\bea
\label{bt}
{N\over 4\pi}\int_\Sigma {\cal A}\wedge \star\, {\cal A}
\eea
and decompose $\cal A$ into harmonic part and fluctuations \cite{Bos:1989wa}
\bea
\label{holA}
{\cal A}={i\pi\over \sqrt{N}} \zeta^{}_I (\Omega_2^{-1})^{IJ} \omega_J^*+{\rm c.c.}+\partial \chi.
\eea
With the boundary term \eqref{bt} $\zeta$ is fixed at the boundary while $\zeta^*$ will be fluctuating freely. 
The wavefunctions of the model, which are holomorphic functions of $\zeta$ \eqref{3dcs} are well known \cite{Bos:1989wa,belov2004conformal,gelca2010classical}. Up to a multiplicative factor due to  small fluctuations, their explicit form  is given by (we only consider trivial spin structure on $\Sigma$)
\bea
\label{3dwavefunction}
    & \Theta_{c_1 \ldots  c_{g}}({ \Omega},\zeta) \ = 
    \ {\rm det}(\Omega_2)\sum\limits_{v_1 \ldots v_g} 
    e^{i \pi\, v^T   { \Omega}v +2\pi i\, v^T   \zeta +\pi\, {\Omega}_2^{-1}\zeta^2/2}.
\eea
Here the sum goes over $v_I=(n_I N+c_I)/\sqrt{N}$, where $n_I\in \Z$ and $c_I\in \Z_N$ parameterize the wavefunction. These wavefunctions are orthogonal with the measure $e^{-{\pi\, \Omega_2^{-1}|\zeta|^2}}$ inherited from \eqref{bt}, and accordingly at the quantum level 
\bea
\zeta^* \rightarrow {\Omega_2\over \pi}{\partial \over \partial \zeta}.
\eea

The wavefunctions \eqref{3dwavefunction} form a representation of the group of Wilson line operators wrapping   cycles $\Gamma(n,m)$ of $\Sigma$, defined as dual to $\Gamma^\vee(n,m) =n^I\omega^{(1)}_I+m^I\omega^{(1)}_{g+I}$,
\bea
W_\Gamma&=&e^{i\oint_\Gamma A}=e^{N^{-1/2}(n+\Omega m)^T{\partial/\partial \zeta}-\pi N^{-1/2} (n+\Omega^* m)^T\Omega_2^{-1}\zeta},\\
W^{}_\Gamma W_{\Gamma'}&=&W_{\Gamma+\Gamma'}e^{{i\pi\over N}(n m'-n'm)}.
\eea
and 
\bea
W_\Gamma \Theta_{c_1 \ldots  c_{g}}(\Omega,\zeta)=\Theta_{c'_1 \ldots  c'_{g}}(\Omega,\zeta) e^{{2\pi i\over N}c\cdot n+{i\pi \over N}n\cdot m},
\eea
where $c'_I=c^{}_I+m_I\, {\rm mod}\, N$.

\subsection{4d geometry}
\label{sec:4dgeometry}
Similarly to 2d, a spin 4-manifold $\M_4$ of zero signature admits a basis or real-valued two-forms with the canonical intersection form 
\bea
\label{intersection}
\int_{\M_4} \omega_i^{(2)} \wedge \omega_j^{(2)}=\eta_{ij},\qquad i,j=1\dots 2n,
\eea
where $\eta$ is defined in \eqref{eta}. The metric 
\bea
\label{4dmetric}
\int_{\M_4}\omega^{(2)}_i \wedge \star\,  \omega^{(2)}_j&=&\g_{ij},\qquad \g=\Lambda^T \Lambda,\qquad 
\Lambda=G^{-1/2}\left(\begin{array}{cc}
B & 1\\
G & 0
\end{array}\right)\\
\g&=&\left(\begin{array}{c|c}
G-BG^{-1}B &-BG^{-1}\\
\hline G^{-1}B & G^{-1}
\end{array}\right),
\eea
can be written in terms of $E=G+B$, the analog of the modular parameter in 2d. In the context of Narain theories positive-definite symmetric $G$ is the metric on the $n$-dimensional torus, while antisymmetric $B$ is the $B$-field. 
The metric \eqref{4dmetric} 
is compatible with $\eta$ in the sense that $\eta^{-1} \g = \g^{-1} \eta$, and hence both can be diagonalized simultaneously in terms of the (anti)-self-dual forms $\star\, \omega_i{\pm}=\pm \omega_i^{\pm}$,  
\bea
\int_{\M_4}\omega^{\pm}_i \wedge \omega^{\pm}_j=\pm \delta_{ij},\qquad i,j,=1\dots n.
\eea
Comparing this with \eqref{2dint} we find that $\omega^{\pm}_i$ are normalized a bit differently, they diagonalize $\g$ while $\omega_I$ does not diagonalize ${\mathbb G}$.
They are related to $\omega^{(2)}_i$ as follows, c.f.~\eqref{real-hol}, 
\bea
\label{real-sd}
\nonumber
\sqrt{2}\,G^{1/2}_{ij}\omega_j^+=\omega^{(2)}_{i}+E_{ij}\, \omega^{(2)}_{n+j},\\
\sqrt{2}\,G^{1/2}_{ij}\omega_j^-=\omega^{(2)}_{i}-E_{ji}\, \omega^{(2)}_{n+j}.
\eea

\subsection{6d geometry and 7d ``Chern-Simons'' theory}
The story in six dimensions is similar to 2d. In what follows we focus on $\M_6=\Sigma\times \M_4$, with $\M_4$ a simply-connected spin 4-manifold of signature zero.
There are $4gn$ real-valued three-forms with the canonical intersection form \eqref{J} of size $4gn$,
\bea
\label{w3}
\omega_A^{(3)}=\{
\omega^{(1)}_I\wedge \eta_{ij}\omega^{(2)}_j,\,  \omega^{(1)}_{g+I}\wedge   \omega^{(2)}_i
\},\qquad I=1\dots g, \quad i,j=1\dots 2n.
\eea
These three-forms are analogs of $\omega_I^{(1)}$ in 2d. 
Next, we introduce a basis of the ``anti-self-dual'' three-forms, the eigenvectors of the Hodge star analogous to $\omega_I$,
\bea
\omega^{}_I \wedge \omega^{+}_i,\qquad \omega^*_I \wedge \omega^{-}_i,\qquad I=1\dots g,\,\, i=1\dots n.
\eea
Comparing with \eqref{real-hol} we find that  correct linear combinations are
\bea\nonumber
\omega_A=\left\{{(G^{-1/2})_{ij}\over \sqrt{2}}(\omega^{}_I\wedge \omega_j^+-\omega^*_I\wedge \omega_j^-),
{(G^{1/2}-BG^{-1/2})_{ij}\, \omega^{}_I\wedge \omega_i^+\over \sqrt{2}}+{(G^{1/2}+BG^{-1/2})_{ij}\, \omega^*_I\wedge \omega_i^-\over \sqrt{2}}\right\},
\eea
such that the relation between $\omega_A$ and $\omega_A^{(3)}$ is given by \eqref{real-hol}
with the 6d ``modular parameter''
\bea
{\mathbf  \Omega}&=&\Omega_1\, \eta^{-1}+ i\, \Omega_2\, \g^{-1},\\
\omega_A&=&\omega_A^{(3)}+{\bf \Omega}_{AB}\omega_{B+{\rm g}}^{(3)},\qquad {\rm g}=2gn.
\eea
The matrix $\bf \Omega$  determines the metric 
\bea
\int_{\M_6}\omega_A^{(3)}\wedge \star\, \omega_B^{(3)}={\bf G}_{AB},\quad {\bf G}={\bf\Lambda}^T {\bf\Lambda},\qquad
{\bf \Lambda}={\bf \Omega}_2^{-1/2}\left(\begin{array}{cc}
 -{\bf \Omega}_1 & 1\\
{\bf \Omega_2} & 0
\end{array}\right),
\eea which is a straightforward generalization of \eqref{2dmetric}.

To quantize the 7d theory \eqref{7d} we add the boundary term \eqref{boundaryterm} and introduce  holomorphic variables $\zeta_A$ via
\bea
\label{H3para}
H_3={i\pi\over \sqrt{N}} \sum_A \zeta_A  ({\bf \Omega}_2^{-1})^{AB}\omega^{*}_B +{\rm c.c.}+\partial \chi.  
\eea
Then the wavefunction (up to contribution of the fluctuating modes) is given by \eqref{3dwavefunction},
\begin{eqnarray}
\label{7dtheta}
     \Theta_{c_1 \dots  c_{\rm g}}({\bf \Omega},\zeta) \ = \ {\rm det}({\bf \Omega}_2)\sum_{v_1  \ldots   v_{\rm g}} e^{i \pi\, v^T   {\bf \Omega}\, v +2\pi i\, v^T   \zeta +\pi\, {\bf\Omega}_2^{-1}\zeta^2/2},
\end{eqnarray}
where $v_A=(N n_A+c_A)/\sqrt{N}$ and the sum is over all $n_A\in \Z$.

To obtain the wavefunctions of section \ref{sec:holquantization} we need to change the variables from $\zeta$ defined in \eqref{H3para} to $\xi,\bar\xi$ defined in \eqref{H3parametrization},
\bea
P\zeta=\left(\begin{array}{c}
\xi\\
\bar \xi\end{array}\right),\qquad P={G^{-1/2}\over \sqrt{2}}\left(\begin{array}{c|c}
G+B & 1\\
\hline
-G+B & 1\end{array}\right).
\eea
Here  $\zeta$ is a $2n$ by $g$ matrix while $\xi,\bar\xi$ are matrices $n$  by $g$. It is then straightforward to check that  $P^TP=\g$, ${\mathcal O}=\Lambda\,\eta$, $u_I={\mathcal O}v_I$ and \eqref{7dtheta} reduces to \eqref{3dtheta}.

One can instead reshuffle \eqref{w3} to 
\bea
\tilde{\omega}_A^{(3)}=\{
-J_{IJ}\omega^{(1)}_{J}\wedge \omega^{(2)}_{i},\,  \omega^{(1)}_{I}\wedge   \omega^{(2)}_{n+i}
\},\qquad I,J=1\dots 2g, \quad i=1\dots n,
\eea
with corresponding holomorphic differentials being
\bea
\tilde{\omega}_A=\left\{{i(\Omega_2^{-1})^{IJ}(G^{1/2})_{ij}\over 2\sqrt{2}}(\omega^{}_J\wedge \omega_j^+-\omega^*_J\wedge \omega_j^-),
{i(\Omega_2^{-1})^{JK}(G^{1/2})_{ij}\over 2\sqrt{2}}(\Omega^*_{IJ}\omega^{}_K\wedge \omega_j^+-\Omega_{IJ}\omega^*_K\wedge \omega_j^-)
\right\}\nonumber
\eea
such that 
\bea
\tilde{\bf \Omega}&=&B J^{-1}+i\, G\, \mathbb{G}^{-1},\\
\tilde{\omega}_A&=&\tilde{\omega}_A^{(3)}+\tilde{\bf \Omega}_{AB}\tilde\omega_{B+{\rm g}}^{(3)},\qquad {\rm g}=2gn.
\eea

Comparing \eqref{H3para} with $\bf \Omega$ substituted by $\tilde{\bf \Omega}$ and  \eqref{H3parametrization} we readily find
\bea
\tilde{P}\zeta=\left(\begin{array}{c}
\xi\\
\bar \xi\end{array}\right),\qquad \tilde{P}={G^{-1/2}\over \sqrt{2}}\left(\begin{array}{c|c}
-\Omega\,  & 1\\
\hline
-\Omega^* & 1\end{array}\right).
\eea
In particular $\tilde{P}^\dagger \Omega_2^{-1} \tilde{P}=G^{-1} \mathbb{G}$ and 
\bea
2\tilde{P}^T \left(\begin{array}{cc}
     0 & \Omega_2^{-1} \\
    \Omega_2^{-1}  & 0
\end{array} \right)  \tilde{P}=\tilde{\bf \Omega}_2^{-1}=G^{-1} \mathbb{G},
\eea
such that  the wavefunction \eqref{7dtheta} reduces to \eqref{5dtheta}.

Finally we discuss how to obtain the 3d theory \eqref{AB} and the 5d theory \eqref{5d} directly from the 7d theory \eqref{7d}.
To that end one introduces one-forms $A^i,B^i$ and expand 
\bea
\label{H3d}
H_3=\sum_{i=1}^n A^i \wedge \omega^{(2)}_i+B^i \wedge \omega^{(2)}_{n+i}.
\eea
This yields \eqref{AB} upon the subsitution into \eqref{7d}.
Accordingly, the boundary term \eqref{boundaryterm} will be given by 
\bea
\label{3dbt}
{N\over 4\pi}\int_\Sigma  \g_{ij}\, (A,B)^i\wedge \star\, (A,B)^j ,
\eea
where $(A^i,B^i)$ is a vector with $2n$ components. 

Similarly, the 5d theory \eqref{5d}, with an additional boundary term ${N\over 4\pi}\int C_2\wedge B_2$, will follow from the 7d after the substitution 
\bea
\label{H5d}
H_3=\sum_{I=1}^g B_2^I\, \omega_I^{(1)}+ C_2^I\, \omega_{n+I}^{(1)},
\eea
while the boundary term \eqref{boundaryterm} becomes
\bea
\label{5dboundaryterm}
{N\over 4\pi} \int_{\M_4} \mathbb{G}_{IJ} (B,C)^I  \wedge \star\, (B,C)^J={N\over 4\pi}\int_{\M_4} \left|\Omega_2^{-1/2}(C_2-\Omega\, B_2)\right|^2. 
\eea
Comparing \eqref{H3d} and \eqref{H5d} with \eqref{H3parametrization} we readily find \eqref{ABxibarxi} and \eqref{BCxibarxi}. 

\section{Proof that $L$ is self-dual}
\label{app:L}
Here we show that additive symplectic code $L$ defined in \eqref{defL} is self-dual.  
In what follows it will be convenient to slightly change the notations of section \ref{sec:topologyandcodes} and make the split of codes of length $g$ into $g'=g-\tilde{g}$ and $\tilde{g}$ explicit
\bea
\nonumber
(a_1,\dots,  a_g,b_1,\dots,b_g) \rightarrow 
(a_1,\dots, a_{g'},b_1,\dots, b_{g'}|a_{g'+1},\dots,a_g,b_{g'+1},\dots,b_g).
\eea
With these notations we introduce the following sets $L_0,L_1$:
\bea
(a_1,\dots a_{g'},b_1,\dots b_{g'})\in L_0,\quad {\rm iff}\quad  
(a_1,\dots a_{g'},b_1,\dots b_{g'}|0,\dots, 0)\in \LL_\gamma,\\
(a_1,\dots a_{g'},b_1,\dots b_{g'})\in L_1,\quad {\rm iff}\quad   (a_1,\dots a_{g'},b_1,\dots b_{g'}|*,\dots, *)\in \LL_\gamma.
\eea
Here $*$ means corresponding element could be arbitrary. Clearly both $L_0,L_1$ are closed under addition, hence these are  additive codes. Trivially, $L_0\subset L_1$.
Since $\LL_\gamma$ is symplectic, $L_0\subset L_1^\perp$. In fact any element ${\rm x}\in L_1^\perp$, if completed by zeros to be of the length $g$, will be orthogonal to any element of $\LL_\gamma$. Hence it belongs to $\LL_\gamma$, and consequently ${\rm x}\in L_0$. This means $L_0=L_1^\perp$ and $L_0$ is a symplectic code (i.e.~self-orthogonal with respect to symplectic inner product \eqref{selforth}). 

We similarly introduce $R_0,R_1$,
\bea
(a_1,\dots a_{\tilde{g}},b_1,\dots b_{\tilde{g}})\in R_0,\quad {\rm iff}\quad  
(0,\dots, 0|a_1,\dots a_{\tilde{g}},b_1,\dots b_{\tilde{g}})\in \LL_\gamma,\\
(a_1,\dots a_{\tilde{g}},b_1,\dots b_{\tilde{g}})\in R_1,\quad {\rm iff}\quad   (*,\dots, *|a_1,\dots a_{\tilde{g}},b_1,\dots b_{\tilde{g}})\in \LL_\gamma,
\eea
and conclude that $R_0=R_1^\perp$ is symplectic. 

Finally we introduce $M_0\subset R_0,M_1\subset R_1$ that include all codewords of the form
\bea
(a_1,\dots, a_{\tilde{g}},0,\dots,0)\in M_0 \quad {\rm iff}\quad  
(0,\dots,0|a_1,\dots, a_{\tilde{g}},0,\dots,0)\in \LL_\gamma,\\
(a_1,\dots, a_{\tilde{g}},0,\dots,0)\in M_1 \quad {\rm iff}\quad  
(*,\dots,*|a_1,\dots, a_{\tilde{g}},0,\dots,0)\in \LL_\gamma.
\eea
Now consider an arbitrary element 
\bea
({\rm x}|{\rm y})\in \LL_\gamma.
\eea
It will contribute to the sum in \eqref{resultingstate} if any only if ${\rm y}\in M_1$. Furthermore each ${\rm x}$ can ``pair'' with exactly $|M_0|$ different elements ${\rm y}\in M_1$. Hence $m({\rm x})=|M_0|$ for any $\rm x$. 

Additive codes $M_0,M_1$ can be defined as 
\bea
M_0=\tilde{\LL}_0\cap R_0,\qquad M_1=\tilde{\LL}_0\cap R_1,
\eea
where $\tilde{\LL}_0$ is the self-dual symplectic code that defines $|0\rangle^{\tilde g}$,
\bea
(a_1,\dots,a_{\tilde{g}},0\dots,0)\in \tilde{\LL}_0,\qquad a_i\in \Z_N.
\eea
Let us now consider a bilinear form defined in terms of the canonical symplectic product on $(\Z_N\times \Z_N)^{\tilde{g}}$,
\bea
({\rm y}_1,{\rm y}_0),
\eea 
where ${\rm y}_1\in R_1, {\rm y}_0\in \tilde{\LL}_0$. To calculate the rank of this form, we note that iff ${\rm y_1}\in \tilde{\LL}_0^\perp=\tilde{\LL}_0$ the bilinear product is zero for any ${\rm y}_0\in \tilde{\LL}_0$, and therefore the rank is $r=|R_1|/|M_1|$. Alternatively, iff ${\rm y}_0\in R_1^\perp=R_0$ the bilinear product is zero for any ${\rm y}_1\in R_1$. As a result 
\bea
\label{rank}
r={|R_1|\over |M_1|}={|\tilde{\LL}_0|\over |M_0|}.
\eea

Our goal now is to calculate the size of $L$. It consists of all distinct ${\rm x}\in L_1$ such that there are ${\rm y}\in M_1$ and $({\rm x}|{\rm y})\in \LL_\gamma$. For each element of $M_1$ there are $|L_0|$ different elements ${\rm x}$ (related by a shift by an arbitrary element from $L_0$), but different ${\rm y},{\rm y}'\in M_1$ related by a shift ${\rm y}-{\rm y}'\in M_0$ will yield the same ${\rm x}$. Hence 
\bea
|L|={|L_0||M_1|\over |M_0|}.
\eea
Taking into account \eqref{rank} and that $|\tilde{\LL}_0|=N^{\tilde{g}}$ we find
\bea
\label{Lsizeprel}
|L|={|L_0||R_1|\over N^{\tilde{g}}}.
\eea

Now we go back to $\LL_\gamma$ and readily conclude that for each ${\rm x}\in L_1$ there are $|R_0|$ possible ${\rm y}$ (all related by shifts from $R_0$) such that $({\rm x},{\rm y})\in \LL_\gamma$. Similarly, for any ${\rm y}\in R_0$ there are $|L_0|$ different ${\rm x}$ such that $({\rm x},{\rm y})\in \LL_\gamma$. And therefore 
\bea
|L_1||R_0|=|L_0||R_1|=|\LL_\gamma|=N^g.
\eea
Together with \eqref{Lsizeprel} this means $|L|=N^{g-\tilde{g}}$, which means $L$ is a symplectic code of maximal size, i.e.~it is self-dual.


\section{Codes over $\Z_N$ for $N=p$ and $N=p^2$}
\label{app:p2counting}
\subsection{Counting ``orthogonal'' codes over $\Z_N\times \Z_N$}
Let us first calculate the number of codes over $\Z_p\times \Z_p$  (for a prime $p$) of length $n$ and with $k$ generators, which are even  in the sense of \eqref{even}. 
For $k=0$ there is a unique such code, consisting of the zero codeword.  For $k=1$ the number is determined by the total number of even non-zero codewords of length $n$, denoted ${\cal N}(n)$ and calculated in \cite{Dymarsky:2025agh} see Appendix F there, 
\bea
{\cal N}(n)=(p^n-1)(p^{n-1}+1)+1.
\eea
Accordingly the number of codes with $k=1$ generator is 
\bea
\label{k=1}
N(k=1,n,p)={{\cal N}(n)-1\over p-1}={(p^n-1)(p^{n-1}+1)\over p-1},
\eea
where the denominator takes into account that the collinear vectors would generate the same code.

For $k=2$ we first can choose one of ${\cal N}(n)-1$ non-zero even vectors, and then supplement it by one of $P(n)-p+1$ even vectors, that are orthogonal to the first one (and not collinear with the first one). Here $P(n)$ is the number of even non-zero vectors orthogonal to any given non-zero even codeword, see the Appendix F of \cite{Dymarsky:2025agh}, 
\bea
P(n)-p+1=(p^n-p)(p^{n-2}+1).
\eea
We therefore find for 
\bea
\label{k=2}
N(k=2,n,p)={({\cal N}(n)-1)(P(n)-p+1)\over (p-1)p(p^2-1)}.
\eea
The denominator here is exactly the size of $GL(2,\Z_p)$ which counts the number of ways the same code can be generated by pairs of different generators.  

Generalizing this to arbitrary $k$ we find  the following formula for the number of even codes of length $n$ and exactly $k$ generators 
\bea
N(k,n,p)=\prod_{i=1}^k {(p^{n-i}+1)(p^{n+1-i}-1)\over p^i-1}.
\eea
This formula agrees with the total number of even self-dual codes $\CC$ when $k=n$ \cite{Aharony:2023zit}
\bea
\label{noc}
\NC(n,p)=N(n,n,p)={|O(n,n,\Z_p)|\over |\Gamma_0|}=\prod_{i=0}^{n-1}(p^i+1).
\eea 
When $N$ is square-free $N=\prod_k p_k$, the total number of codes is simply the product of \eqref{noc}
\bea
\NC(n,N)=\prod_k \NC(n,p_k).
\eea 

Now we are ready to calculate the number of even self-dual codes of length $n$ over  $\Z_N \times \Z_N$ where $N=p^2$ and $p$ is a prime. Such codes  fall into different orbits $O_a$ based on their group structure, $\Z_N^{a} (\Z_p \times \Z_p)^{(n-a)}$, parameterized by $a=0\dots n$. To calculate the number of codes in each orbit we can consider a map mod $p$ such that an even self-dual code of type $\Z_N^{a} (\Z_p \times \Z_p)^{n-a}$ will become an even code over $\Z_p \times \Z_p$ with $k=a$ generators. There are $N(k,n,p)$ codes of this type. Besides, the map has a non-trivial kernel. It is clear that if we consider a code with the group structure $\Z_N^m$ of the form $(a_1\dots a_m,0\dots 0)$ for $a_i\in \Z_N$, then any orthogonal transformation of the form 
\begin{eqnarray}
    \left(\begin{array}{cc}
    1 & X\\
    0 & 1
    \end{array}\right)
\end{eqnarray}
with any antisymmetric matrix $X=0\, {\rm mod}\, p$ will not change the resulting code over $\Z_p$. There  are $p^{a(a-1)/2}$ such matrices $X$.
This fully describes the degeneracy of the mod $p$ map for this given code. The size of the kernel mapping even self-dual codes over   $\Z_N \times \Z_N$ to even codes over $\Z_p \times \Z_p$ is the same for all codes with the same group structure.  Hence we find for the number of codes of type $\Z_N^{a} (\Z_p \times \Z_p)^{(n-a)}$ is 
\bea
|O_a|\equiv \NC(a,n,p^2)=p^{a(a-1)/2} N(a,n,p).
\eea
This formula matches previously known results for particular $p$ and $n$. 
The total number of even self-dual codes over $\Z_N\times \Z_N$ for $N=p^2$ is 
\bea
\NC(n,p^2)=\sum_{a=0}^n \NC(a,n,p^2).
\eea

\subsection{Counting ``symplectic'' codes over $\Z_N\times \Z_N$}
We start by counting the number of   codes  over $\Z_p\times \Z_p$ of length $g$ symplectic in the sense of \eqref{selforth} with exactly $k$ generators. For $k=1$ there are 
\bea
N_s(k=1,g,p)={p^{2g}-1\over p-1}
\eea
such codes, where the numerator evaluates the number of non-zero vectors of length $2n$ and the denominator is the same as in \eqref{k=1}.
For $k=2$, there are 
\bea
N_s(k=2,g,p)={(p^{2g}-1)(p^{2g-2}-1)p\over (p-1)p(p^2-1)}
\eea
such codes etc. In the end we find for arbitrary $k$
\bea
N_s(k,g,p)=\prod_{i=1}^k {(p^{2(g+1-i)}-1)\over (p^i-1)}.
\eea
When $k=g$ the number of self-dual symplectic codes $\LL$ is \cite{Dymarsky:2025agh}
\bea
\NL(g,p)=N_s(g,g,p)={|Sp(2g,\Z_p)|\over |\Gamma_0|}=\prod_{i=1}^g (p^i+1).
\eea
Again, when $N=\prod_k p_k$ is square-free, the total number of codes is
\bea
\NL(n,N)=\prod_k \NL(n,p_k).
\eea 

To calculate the number of symplectic codes over $\Z_N\times \Z_N$ for $N=p^2$, we notice that such codes split into orbits $S_a$ based on their group structure  $\Z_N^{a} (\Z_p \times \Z_p)^{(g-a)}$, 
parameterized by $a=0\dots g$. To calculate the number of codes in each orbit we can consider a map mod $p$ such that a symplectic self-dual code of type $\Z_N^{a} (\Z_p \times \Z_p)^{g-a}$ will become symplectic code over $\Z_p \times \Z_p$ with $k=a$ generators. There are $N_s(k,n,p)$ codes of this type. Besides, the map has a non-trivial kernel. It is clear that if we consider a code with the group structure $\Z_N^m$ of the form $(a_1\dots a_m,0\dots 0)$ for $a_i\in \Z_N$, then any orthogonal transformation of the form 
\begin{eqnarray}
    \left(\begin{array}{cc}
    1 & X\\
    0 & 1
    \end{array}\right)
\end{eqnarray}
with any symmetric matrix $X=0\, {\rm mod}\, p$ will not change the resulting code over $\Z_p$. There  are $p^{a(a+1)/2}$ such matrices $X$.
This fully describes the degeneracy of the mod $p$ map for this given code. The size of the kernel mapping symplectic self-dual codes over   $\Z_N \times \Z_N$ to symplectic codes over $\Z_p \times \Z_p$ is the same for all codes with the same group structure.  Hence we find for the number of codes of type $\Z_N^{a} (\Z_p \times \Z_p)^{(g-a)}$ is 
\bea
|S_a|\equiv \NL(a,g,p^2)=p^{a(a+1)/2} N_s(a,g,p).
\eea
As a consistency check we note that 
\bea
|S_g|={|\Sp(2g,\Z_p^2)|\over |\Gamma_0|}.
\eea
The total number of even self-dual codes over $\Z_N\times \Z_N$ for $N=p^2$ is 
\bea
\NL(g,p^2)=\sum_{a=0}^n \NL(a,g,p^2).
\eea

\subsection{Sum over topologies for $g=1$}
\label{matchingRHS}
In what follows we focus on the case $g=1$ and evaluate the sum over 3d topologies ending on a torus for the AB theory \eqref{AB} with $N=p^2$ by {\it matching} the LHS of \eqref{3dholographyp2}.
In other words, instead of performing a genuine sum using genus reduction, we note that the resulting sum at genus $g=1$ includes only states $|\LL\rangle$ \eqref{Lstate} for the symplectic self-dual codes $\LL$ of length $g=1$ over $\Z_N\times \Z_N$, $N=p^2$, and match corresponding coefficients. We leave the task of evaluating  the  sum over 3d manifolds from the first principles  for the future. 

Symplectic codes of length $g=1$ over $\Z_N\times \Z_N$ with $N=p^2$ split under the action of $SL(2,\Z)$  into two orbits, $S_1$ and $S_0$. First has the size $|S_1|=p(p+1)$ and includes codes of the form  
$(a,r\, a)$ for any $a\in \Z_N$ and $r=0\dots N-1$, as well as $(rpa,a)$ where $r=0\dots p-1$. 
The orbit $S_0$ includes just a single code of the form $\LL'=(pa,pb)$, where $a,b\in \Z_p$. Obviously this code is invariant under $SL(2,\Z)$ by itself. 

We want to find coefficients $\alpha,\beta$ such that 
\bea
\nonumber
\sum_{a=0}^n \sum_{{\cal C}\in O_a} |{\cal C}\rangle ={ \alpha\over N^{n/2}} \sum_{\LL\in S_1}|\LL\rangle+{\beta\over N^{n/2}}\sum_{\LL\in S_0}|\LL\rangle ={\alpha} \sum_{\gamma \in \Gamma_0\backslash SL(2,\Z_N)} U_\gamma |0\rangle_{3d} +{\beta} |p\rangle.\\
\label{Hol}
\eea
Here we used that the vacuum state of the 3d theory \eqref{AB} on a torus
\bea
\label{vacuum}
|0\rangle_{3d}={1\over p^n}\sum_{a_i\in \Z_k} |a_1,0,\dots a_n,0\rangle_{5d}
\eea
 is the $n$-th tensor power of the stabilizer state $|\LL_0\rangle=\sum_{{\rm c}\in \LL_0}|{\rm c}\rangle_{5d}$ for $\LL_0=(a,0)$ up to an overall normalization factor $N^{n/2}$. 
We also introduced the state 
\bea
\label{L'}
|p\rangle={1\over N^{n/2}} |\LL'\rangle= {1\over p^n}\sum_{a_i,b_i\in \Z_p} |p a_1,p b_1,\dots p a_n,p b_n\rangle_{5d}
\eea
such that $\langle p |p\rangle=1$. 

To calculate the coefficients $\alpha,\beta$ we can evaluate them first for each $O_a$:
\bea
\label{Hk}
{1\over \NC (a,n,p^2)}\sum_{{\cal C}\in O_a} |\CC\rangle =\alpha_a \sum_{\gamma \in \Gamma_0\backslash SL(2,\Z_N)} U_\gamma |0\rangle +\beta_a |p\rangle.
\eea
We obtain the first equation for $\alpha_a,\beta_a$ by evaluating the scalar product of \eqref{Hk} with $\langle 0|$:
\bea
\label{C1}
1=\alpha_a \left(1+{(p-1)\over p^n}+{p^2\over p^{2n}}\right) +{\beta_a\over p^n}
\eea
where $\langle 0|p\rangle=1/p^n$ follows from \eqref{vacuum} and \eqref{L'} and we used explicit expression for all $p^2+p$ codes in the orbit of $(a,0)$ to evaluate their scalar product with $(a,0)$. 

To obtain second equation, we evaluate the scalar product with $\langle p|$. In this case the RHS is simple. Since $\langle p|$ is modular-invariant all $p^2+p$ states in the $SL(2,Z)$ orbit of $|0\rangle$ have the same scalar product. Evaluation of the LHS can proceed as follows. 
All codes ${\cal C}$ in $O_a$ are mapped to each other by the orthogonal group. State $|p\rangle$ is invariant under this group, hence we can evaluate the scalar product between $|\LL'\rangle$ and $|\CC\rangle$ for any chosen $\CC\in S_0$. It is convenient to choose it in the form 
\bea
{\cal C}=(\alpha_1,0,\dots \alpha_{a},0,p\alpha_{a+1},p\beta_{a+1},\dots p\alpha_n,p\beta_n).
\eea
Here $\alpha_i\in \Z_N$ for $i\leq n-a$ and $\alpha_i,\beta_i\in \Z_p$ for $i>n-a$.
We stress, these are orthogonal codes, not symplectic ones. A beautiful thing is that the code ${\cal L'}$ is both symplectic and orthogonal. Hence the scalar product between corresponding states is 
\bea
{\langle \LL'|\CC\rangle\over N^n}={p^{2n-a}\over p^{2n}}.
\eea
Combining all together we find
\bea
\label{C2}
{p^{n-a}}=\alpha_a {p^2+p\over p^n}+\beta_a,
\eea
and 
\bea
\label{largep}
\alpha_a&=&\frac{\left(p^a-1\right) p^{2 n-a}}{\left(p^n-1\right) \left(p^n+p\right)},\\
\beta_a&=&\frac{p^{n-a} \left(p \left(p-(p+1) p^a\right)+p^{2 n}+(p-1) p^n\right)}{\left(p^n-1\right) \left(p^n+p\right)}.
\eea
Eventually we get for the coefficients in \eqref{Hol} and \eqref{3dholographyp2}
\bea
\label{al}
A_{p^2}={\alpha\over p^n \NC(n,p^2)},\qquad \alpha&=&\sum_{a=0}^n \alpha_a  |O_a|,\\
B_{p^2}={\beta\over p^n \NC(n,p^2)},\qquad \beta&=&\sum_{a=0}^n \beta_a  |O_a|. \label{bet}
\eea

In the large $p\rightarrow \infty$ limit we find for $n>2$,
\bea
\alpha_a=1-\delta_{a,0},\qquad \beta_a=p^{n-a},\quad |O_a|\approx (1+\delta_{a,n})p^{a(2n-1-a)},
\eea
and the sum in \eqref{bet} is saturated for $a=n-1$,
\bea
\label{3dp2}
{1\over \NC}\sum_{\CC} |\CC\rangle= \sum_{\gamma \in \Gamma_0\backslash SL(2,\Z_{p^2})} U_\gamma |0\rangle +{p^{1-n}\over 3} |\LL'\rangle,\qquad \NC\approx 3 p^{n(n-1)}.
\eea
We note that $|\LL'\rangle^{\otimes n}=|\CC'\rangle^{\otimes g}$ where $\CC'\in O_0$ is the unique code in the orbit $O_0$ of the form $(p\alpha,p\beta)\in \CC'$, $\alpha,\beta\in \Z_p^n$.

The expression \eqref{3dp2} is valid also for finite $p$ and $n\gg 1$. It is interesting to note that in the large central charge limit $n\rightarrow \infty$ only handlebody contributions survive, confirming the expectation of \cite{Nick}.

In the reminder of this section we switch to  5d theory (still for $g=1$) and evaluate the sum over 5d topologies by matching the LHS of \eqref{holid5d}, 
\bea
\sum_{\LL\in S_0}|\LL\rangle+\sum_{\LL\in S_1}|\LL\rangle=\sum_{\CC \in O_a} {\delta_a\over |O_a|} |\CC\rangle.
\eea
On the LHS all codes (Maxwell theories) enter with the same coefficient. On the RHS coefficients are ambiguous because different states $|\CC\rangle$ are linearly-dependent. Terms on the RHS have the interpretation of 5d topologies, but there is no ``first principles'' way to fix $\delta_a$ because of this ambiguity. Similarly to  \eqref{C1} and \eqref{C2} we obtain 
\bea
\sum_{a=0}^n \delta_a=p^n\left(1+{p-1\over p^n}+{p^2\over p^{2n}}+{1\over p^n}\right),\\
\sum_{a=0}^n \delta_a p^{n-a}=p^n\left({p^2+p\over p^n}+1\right).
\eea
For $g=1$, there are two  states  invariant under both orthogonal and symplectic groups, hence any two orbits will suffice. 
We find most convenient to keep only $a=n$ and $a=0$, while all other $\delta_k$ are taken to be zero,
\bea
&&A'_{p^2}={\delta_n\over |O_n|\NL},\quad \delta_n=p^n+p-p^{2-n},\qquad \, \delta_1=\dots=\delta_{n-1}=0, \\
&&B'_{p^2}={\delta_0\over \NL},\qquad\quad  \delta_0=p^{2-n}+p^{2(1-n)},\qquad \NL=p^2+p+1.
\eea
In the large $p$ limit and $n>2$ we find 
\bea
\delta_0=p^{2-n},\quad \delta_n=p^n,
\eea
and
\bea
{1\over \NL} \sum_{\LL} |\LL\rangle ={p^{2(n-1)}\over |O_n|}\sum_{h\in \Gamma_0\backslash O(n,n,\Z_{p^2})} U_h |0\rangle_{5d} + p^{-n}|\CC'\rangle.
\eea
When $n=1$ we find for arbitrary $p$, 
\bea
\label{g=n=1}
{1\over \NL} \sum_{\LL} |\LL\rangle ={p^2\over p^2+p+1}\sum_{h\in \Gamma_0\backslash O(n,n,\Z_{p^2})} U_h |0\rangle_{5d} + {p\over p^2+p+1}|\CC'\rangle.
\eea

\acknowledgments
We thank Ofer Aharony and Max H$\ddot{\rm u}$bner for correspondence and  Igor Klebanov and Juan Maldacena for reading the manuscript. 

A.D.~is grateful to the International Centre for Mathematical Sciences for its hospitality and the opportunity to conduct research at the James Clerk Maxwell Birthplace. 
A.D.~acknowledges support from the IBM Einstein Fellow Fund and NSF under grant 2310426.



\bibliographystyle{JHEP}
\bibliography{biblio.bib}


\end{document}